\documentclass{aa}  

\usepackage{graphicx}
\usepackage{xcolor}
\usepackage{txfonts}
\usepackage{longtable}
\usepackage[T1]{fontenc}
\usepackage{subfigure,wrapfig}
\usepackage{soul}
\usepackage[normalem]{ulem}
\usepackage[switch]{lineno}
\usepackage{changes}
\usepackage{hyperref}
\begin{document}

   \title{Mass and wind luminosity of young Galactic open clusters in Gaia DR2}


   \author{S. Celli
          \inst{1,2}
          \and
          A. Specovius
          \inst{3}
          \and
          S. Menchiari
          \inst{4}
          \and
          A. Mitchell
          \inst{3}
          \and 
          G. Morlino
          \inst{4}
          }

   \institute{Sapienza Universit\`a di Roma, Physics Department,
                 P.le Aldo Moro 5, 00185, Rome, Italy\\
                 \email{silvia.celli@uniroma1.it}
         \and
                 Istituto Nazionale di Fisica Nucleare, Sezione di Roma,
                 P.le Aldo Moro 5, 00185, Rome, Italy
         \and
		 Friedrich-Alexander Universit\"at Erlangen-N\"urnberg, Erlangen Centre for Astroparticle Physics,
		 Nikolaus-Fiebiger-Str. 2, 91058 Erlangen, Germany
	 \and
                Istituto Nazionale di Astrofisica, Osservatorio Astrofisico di Arcetri,
                L.go E. Fermi 5, Firenze, Italy
   }


 
  \abstract
   {Star clusters constitute a significant part of the stellar population in our Galaxy. The feedback processes they exert on the interstellar medium impact multiple physical processes from the chemical to the dynamical evolution of the Galaxy. In addition, young and massive stellar clusters might act as efficient particle accelerators and contribute to the production of cosmic rays.}
   {We aim at evaluating the wind luminosity driven by the young (< 30 Myr) Galactic open stellar clusters observed by the Gaia space mission. This is crucial for determining the energy channeled into accelerated particles.}
   {To do this, we developed a method relying on the number, magnitude, and line-of-sight extinction of the stars observed per cluster. Assuming that the stellar mass function follows a Kroupa mass distribution and accounting for the maximum stellar mass allowed by the age and mass of the parent cluster, we conservatively estimated the mass and wind luminosity of 387 local clusters within the second data release of Gaia.}
   {We compared the results of our computation with recent estimates of young cluster masses. With respect to these, our sample is three times more abundant, particularly above a few thousand solar masses. This is of the utmost relevance for predicting the gamma-ray emission resulting from the interaction of accelerated particles. The cluster wind luminosity distribution we obtained extends up to $3 \times 10^{38}$~erg/s. This is a promising feature in terms of potential particle acceleration scenarios.}
    {} 


   \maketitle
%

\section{Introduction}
Stellar clusters belonging to the disk of the Galaxy are usually referred to as open clusters. They consist of homogeneous groups of stars with the same age and initial chemical composition. As star formation occurs primarily in clusters, they constitute ideal laboratories for tracing the star formation and evolution in our Galaxy \cite[see][for a recent review]{krumholz2019}. The dynamical evolution of clusters is a result of several processes, including stellar evolution, two-body relaxation, tidal interactions, and shocks. During their lifetime, clusters tidally interact with their parent giant molecular clouds and with the Galactic structure. They eventually dissolve after a few relaxation times. 

Young clusters, in particular, are key to understanding star formation processes because they  enable the initial mass function (IMF) to be traced. The most massive of these objects, so-called young and massive stellar clusters (YMSCs), have recently received attention in the high-energy astroparticle physics community because they have been detected in very high-energy (VHE, $\gtrsim100$\,GeV) gamma rays \citep{felixCluster} with the High Energy Stereoscopic System (H.E.S.S.). Morphological measurements of this emission have enabled  the derivation of the spatial distribution of accelerated particles in the surroundings of several Galactic clusters (including Westerlund 1 and the Cygnus Cocoon). The gamma-ray profiles and intensities suggest a continuous injection that may be powered by the stellar winds of massive stars provided that $\sim 1\%$ of the wind power is converted into accelerated particles. The Cygnus Cocoon has even been observed at the highest energies ever achieved in gamma-ray astronomy: two photons with an energy above 1 PeV (1 PeV = $10^{15}$~eV) have been detected from the Large High Altitude Air Shower Observatory (LHAASO) \citep{lhaaso2023-Cyg}. These observations probed the capabilities of stellar clusters to operate as PeV-particle accelerators (i.e., PeVatrons) and beyond. 

Explaining the origin of cosmic rays (CRs) in the PeV domain remains one of the challenges in astroparticle physics today. While supernova remnants (SNRs) are often addressed to explain their energetics, the maximum energy that can be produced by these sources remains unclear \citep{celli2019, Cristofari+2020, Diesing:2023}. On the other hand, the power provided by stellar winds can be $\gtrsim 10\%$ of the power provided by SNRs \citep{Cesarsky-Montmerle:1983, Morlino-ICRC:2023}. Moreover, YMSCs have favorable conditions in terms of shock size and duration, so that it is plausible that energies up to $\sim$PeV can be obtained \citep{morlino2021}.
For these reasons, YMSCs were suggested as candidate sources of Galactic CRs up the \emph{knee} region, complementary to SNRs. In addition to energetic considerations, acceleration of particles from the wind of massive stars appears to be a necessary component to explain the $^{22}$Ne/$^{20}$Ne anomaly in CR composition \citep{neon, Prantzos2012, Tatischeff+2021}: the enhancement observed with respect to solar abundances requires an acceleration from a carbon-enriched medium rather than from the standard interstellar medium (ISM), such as from the winds of massive stars. The relative contribution of these different source populations, SNRs and YMSCs, to the observed CRs is yet to be clarified, as is their role in energy, and in particular, in the region of about a few PeVs \citep{vieu2023, Morlino-ICRC:2023}. 

Furthermore, the exact location where particle acceleration would take place in YMSCs is a topic of active debate in the community. Several proposals have been put forward, based either on the acceleration in wind-wind collision regions \citep{Bykov+2020}, on stochastic second-order acceleration in the highly turbulent medium of the cluster core \citep{klepach2000}, or on acceleration at the wind termination shock (WTS) \citep{morlino2021}. When a cluster is older than a few million years, supernova (SN) explosions start to dominate the dynamics \citep{vieu2022}, but for younger clusters, the energy input available for particle acceleration can only come from stellar winds. Understanding their exact kinetic luminosity is therefore of paramount importance. It is therefore necessary to characterize the cluster physical properties, in particular, their mass and wind luminosity. This is the objective of this work, while in two accompanying papers, we will explore the acceleration model and its radiative signatures. In all of the three works, we focus on young clusters for the reasons explained above. 

The Gaia satellite has recently performed the most accurate astrometric and photometric survey of the Milky Way. This extended our knowledge about the Galaxy structure and evolution. The application of advanced classification algorithms based on machine-learning techniques to its second data release (DR2) has resulted in the identification of an extended sample of clusters \citep{gaia2020}. Even though this catalog does not correspond to the latest available data release, it nevertheless contains detailed information concerning cluster member stars and their physical parameters, which are crucial (in the context of the WTS particle acceleration model) to evaluate the luminosity of the collective wind.

This paper is organized as follows. In Sec.~\ref{sec:data}, we introduce and characterize the properties of the open cluster data sample we adopted for this study, which is devoted to systems younger than 30~Myr. We aim to provide all the necessary ingredients for the computation of the particle acceleration spectrum produced at the cluster WTS. We therefore describe the method we developed for estimating the mass and wind power of each cluster, for which we relied on the properties of the observed cluster star members reported by Gaia. In particular, we obtain in Sec.~\ref{sec:stars} a lower limit to the values of cluster masses by assuming the same stellar mass distribution function for all young clusters and normalizing it to reproduce the Gaia observations in terms of the number of stars within a given magnitude range. We further account for the optical extinction in the cluster direction. We compare this result with existing cluster mass estimates in Sec.~\ref{sec:comparison}. A fraction of the cluster sample investigated here was also reported in the Milky Way Stellar Cluster (MWSC) catalog and mass values derived by \citet{just2023} through their tidal parameters. Moreover, the recent study by \citet{almeida2023} yielded mass values for some of the clusters belonging to the same Gaia data release we investigated from an approach based on a Monte Carlo method. We hence discuss the compatibility of the different results. In Sec.~\ref{sec:wr}, we  present a procedure to evaluate the related mass uncertainties based on the number of observed WR stars in Galactic open clusters. By including the contribution of observed WRs, we further derive the cluster wind luminosities in Sec.~\ref{sec:lum}. 
Finally, we conclude in Sec.~\ref{sec:concl} and additionally provide three appendices that discuss technical aspects of our cluster mass estimation approach, the energetic contribution of SNe occurring in the cluster sample under investigation, and details about the online supplementary data file.

\section{Data sample and methods}
\label{sec:data}
In its DR2, 2017 open clusters have been identified in the Gaia data \citep{gaia2020}. For 1867 of these, cluster parameters such as the distance modulus, extinction, and age were obtained through artificial neural network techniques. Compared to previous 
surveys, such as the MWSC catalog \citep{mwsc2013}, the sample of open clusters from Gaia DR2 reports detailed 3D information that is only allowed by astrometric measurements (proper motions and parallaxes). Therefore, we decided to adopt the Gaia DR2 open cluster catalog as a reference for our study, and to complete missing information about clusters that is required by the physical modeling from the MWSC survey. 

An important feature of the Gaia release is that it provides values for the cluster age. This is a crucial parameter for the application of acceleration models because the main source of energy in these systems changes as they evolve. The energetics of star clusters is dominated by stellar winds for the first few million years and subsequently by SNR shocks \citep{vieu2022}. Therefore, we applied a selection criterion to the initial cluster sample and only retained those younger than 30\,Myr, which roughly corresponds to the main-sequence (MS) lifetime of stars with $M\simeq 8\rm M_{\odot}$, that is, the minimum mass value for stellar evolution to terminate with an SN explosion. This reduced the number of considered clusters to 390. 

For each cluster in the selected sample, we derived the wind luminosity, which strongly depends on the cluster evolutionary stage and composition in terms of its member stars. However, because of the limitations due to instrumental sensitivity and light extinction, the  actual stellar population of each cluster had to be inferred from the observed one.  

Our approach is based on the assumption of a stellar mass distribution inside the cluster, $\xi(M) \equiv dN/dM$, for which we derived the correct normalization to reproduce the number of detected stars per cluster, $N^*$, in the given magnitude range of the observations (for further details, see Sec.~\ref{sec:stars}). To do this, 
\begin{itemize}
\item[i)] we first extracted the number and magnitude of the observed member stars for every young cluster reported by \citet{gaia2020}, as well as the extinction toward the corresponding direction;
\item[ii)] we then obtained the observed minimum and maximum G-band magnitude of its member stars through a kernel density fit of the stellar magnitude distribution of each cluster;
\item[iii)] we converted these values into the maximum and minimum intrinsic luminosity of member stars, respectively, through both bolometric and extinction corrections;
\item[iv)] we later assumed the mass-luminosity relation of MS stars to derive the expected maximum and minimum stellar mass observed per cluster, which we labeled $M^*_{\rm max}$ and $M^*_{\rm min}$, respectively;
\item[v)] we finally derived the normalization of the  stellar mass function for each cluster to reproduce the observed number of stars $N^*$ in the derived mass range $[M^*_{\rm min}; M^*_{\rm max}].$
\end{itemize}

\begin{figure}
\centering
\subfigure[\label{fig:age}]{\includegraphics[width=0.47\textwidth]{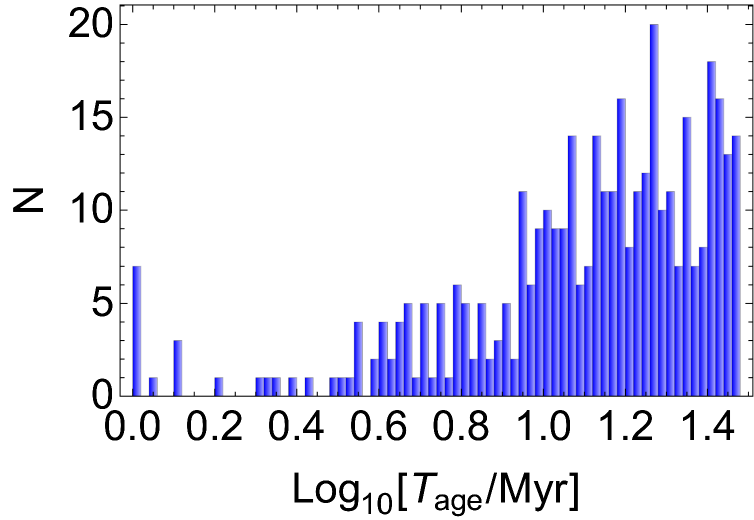}}
\subfigure[\label{fig:distance}]{\includegraphics[width=0.47\textwidth]{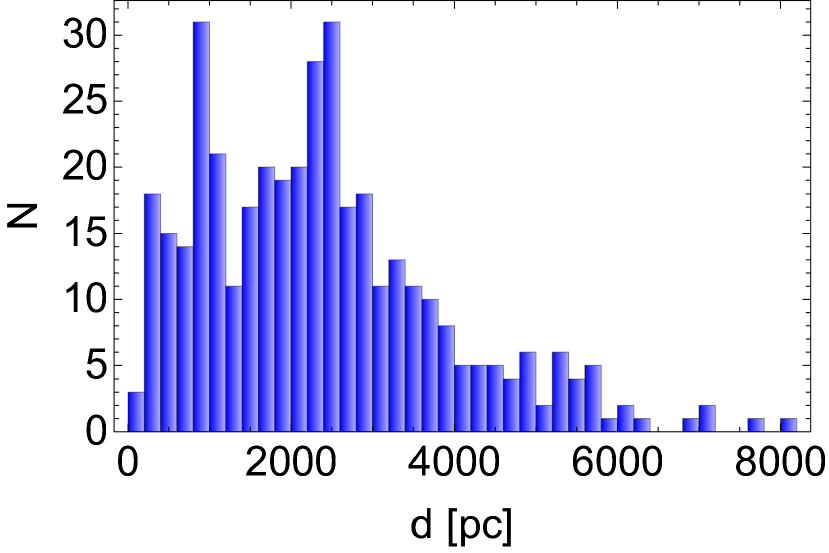}}
\subfigure[\label{fig:nstar}]{\includegraphics[width=0.47\textwidth]{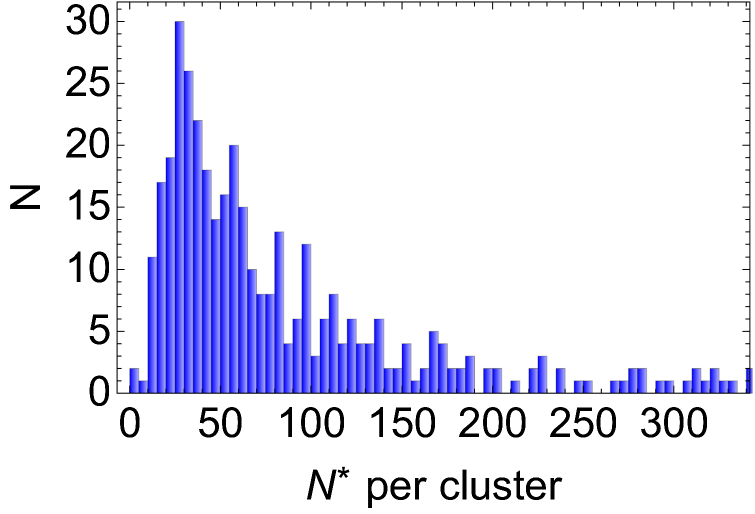}}
\caption{Distributions of physical parameters regarding the selected cluster sample from Gaia DR2 (i.e. younger than 30~Myr). (a) Cluster ages in units of Myr. (b) Cluster distances in units of pc. (c) Number of observed member stars per cluster.}
\label{fig:clusters}
\end{figure}

Since the correction for G-band extinction is based on the measured stellar effective temperature, the lack of this parameter in three clusters of the sample (i.e., IC~2948, NGC~1333, and Pismis~16) reduced the size of the final cluster sample to 387. The distribution of their age and distance from Earth are shown in Figure~\ref{fig:clusters}, with median values of $14.5$~Myr and $2.3$~kpc, respectively.
The spatial distribution of our sample (panel b) is almost flat up to $\sim 3$\,kpc with large fluctuations, while it decreases at larger distances. This likely indicates that the sample is complete up to this distance. The fluctuations are probably due to the local inhomogeneities.
The actual distance up to which the Gaia sample can be considered complete is not known with accuracy: The analysis of the extinction maps \citep{Lallement+2022} suggests that clusters should be detectable up to 3~kpc, even though there are directions in the Galactic plane where the extinction is high even at closer distances.

Figure~\ref{fig:clusters} also reports the number of stars associated with each cluster, from which we derived the correct normalization of the mass distribution. After this was obtained, we derived the current stellar mass function of each cluster by accounting for stellar luminosities beyond the Gaia sensitivity following theoretical considerations. In particular, we fixed an absolute minimum stellar mass $M_{\rm min}$ to $0.08\,M_\odot$, corresponding to the limiting mass value in brown dwarfs for which nuclear burning is inhibited \cite[see][]{Richer+2006Sci}. The determination of the current maximum stellar mass was more involved. We fixed the absolute maximum stellar mass in the Milky Way to $\sim 150\,\rm M_{\odot}$ as inferred from former analyses of the Arches cluster \citep{Figer:2005, Kroupa:2005Natur}. More recent works investigating the cluster R~136 inferred a maximum stellar mass as high as $\sim 300\, \rm M_{\odot}$ \citep{Crowther+2010}. However, we here adopted the conservative value of 150 M$_{\odot}$, also because higher values do not change our results significantly given the very low number of young and massive star clusters in the sample that might in principle host very massive stars. However, it is reasonable to assume that the maximum stellar mass depends on the total mass of the parent cluster. Even though this connection is still a matter of debate, we relied on the relation found by \citet{Weidner-Kroupa:2006}, where the maximum stellar mass roughly corresponds to $\sim 10\%$ of the cluster mass (see also \citet{Yan-Jerabkova-Kroupa:2023} for a more recent study of the correlation between the maximum stellar mass and the parent cluster mass).

Stellar evolution affects the high-mass tail of the distribution, which becomes progressively depleted because of SN explosions. In other words, the older the cluster, the lower the mass of the currently most massive member star \citep{buzzoni2002}. For the sake of completeness, we mention that the number of massive stars may also be reduced because of dynamical effects that may result in the ejection of stars from the cluster. \citep{Oh-Kroupa-Pflamm:2015} have estimated that clusters more massive than $\sim 10^3\, \rm M_{\odot}$ may eject a fraction of O-type stars up to $\sim 20\%$ with a velocity $\gtrsim 10 \, \rm km/s$. We neglected this complication, which may lower the final estimate of the cluster kinetic luminosity by a few tens of percent. 
For the purposes of our computations, we constrained $M_{\rm max}$ to be the lower of these values. Further details about the specific values obtained per cluster are given in Sec.~\ref{sec:stars}. Because the mass estimation procedure defined here strongly relies on the capabilities of a single star identification, our result should rather be interpreted as a lower bound to the actual cluster mass, and hence mass-loss rate and wind power, as further discussed in the following sections.

\section{Estimating cluster masses}
\label{sec:stars}
The details of the steps undertaken in the mass estimation procedure are explained here. \\

(i) First, we evaluated individual cluster masses by assuming a broken power law functional form with four components for the stellar mass function $\xi(M)$. Following \citet{weidner2004}, we set
\begin{linenomath}
\begin{equation}
 \xi(M) = k \times \left\{
 \begin{array}{l@{\hspace{.3cm} \vspace{0.1cm}}c}
  \left(\frac{M}{M_{\rm min}} \right)^{-\beta_1} & M_{\rm min} \leq M < M_0 \\
  \left(\frac{M_0}{M_{\rm H}}\right)^{-\beta_1} \left(\frac{M}{M_0}\right)^{-\beta_2} & M_0 \leq M < M_1 \\
  \left( \frac{M_0}{M_{\rm H}} \right)^{-\beta_1} \left(\frac{M_1}{M_0}\right)^{-\beta_2} \left(\frac{M}{M_1}\right)^{-\beta_3} & M_1 \leq M < M_{\rm max} \\
 \end{array}
 \right.
\label{eq:imf}
\end{equation}
\end{linenomath}
with $\beta_1=1.30$, $\beta_2=2.30$, $\beta_3=2.35$, $M_{\rm min}=0.08\,M_\odot$,  $M_0=0.50\,M_\odot$, and $M_1=1.00\,M_\odot$, $M_\odot$ being the mass of the Sun.  \\

(ii) Next, we applied to each cluster a kernel fit procedure to the G-band magnitude distribution of all its stellar members. The fit was applied to extract both the minimum $\mathcal{M}^G_{\rm min}$ and the maximum $\mathcal{M}^G_{\rm max}$ magnitudes of stars detected per cluster. In place of the maximum magnitude, we considered the mode of the stellar magnitude distribution per cluster, as it corresponds to the magnitude value up to which the Gaia survey data can be considered complete, as detailed in Appendix~\ref{sec:appA}. \\
Then, we derived the corresponding bolometric magnitudes $\mathcal{M}_\mathrm{bol}$ via the bolometric correction factor $BC_G$. 
For a stellar G-band magnitude $\mathcal{M}_{G}$, this conversion is 
\begin{linenomath}
\begin{equation}
\mathcal{M}_\mathrm{bol} = \mathcal{M}_{G} + BC_G \, ,
\end{equation}
where $BC_G$ is obtained considering the observed effective temperature $T_\mathrm{eff}$ of the stars within the cluster
\begin{equation}
BC_G (T_\mathrm{eff}) = \sum\limits_{i=0}^4 \alpha_i \left(T_\mathrm{eff} - T_{\mathrm{eff},\odot}\right)^i \, ,
\end{equation}
\end{linenomath}

\noindent
with the parameters $\alpha_i$ given by Gaia FLAME\footnote{\url{https://gea.esac.esa.int/archive/documentation/GDR2/Data_analysis/chap_cu8par/sec_cu8par_process/ssec_cu8par_process_flame.html}}. \\

(iii) We converted the apparent bolometric magnitudes into intrinsic magnitudes by extracting line-of-sight extinction values $A_G$ from the v6 Gaia Starhorse catalog\footnote{\url{https://doi.org/10.17876/gaia/dr.2/52}}. From these, the stellar luminosities were computed as
\begin{linenomath}
\begin{equation}
L_s =  10^{0.4(\mathcal{M}_{\odot,\mathrm{bol}} - \mathcal{M}_\mathrm{bol} + A_G)} \, L_{\odot} \,.
\end{equation}
\end{linenomath}

(iv) These luminosities were used to derive the corresponding stellar mass $M_s(L_s)$ by inverting the following mass-luminosity relation, in the form of a broken power law,
\begin{linenomath}
\begin{equation}
\label{eqn:Lstar}
L_{\rm s}(M)=
\begin{cases}
 L_{b1}\left(\frac{M}{M_{b1}} \right)^{\gamma_1} \left[\frac{1}{2}+ \frac{1}{2} \left(\frac{M}{M_{b1}} \right)^{1/\Delta_1} \right]^{\Delta_1(-\gamma_1+\gamma_2)} & M/M_{\odot} < 12 \\
 \epsilon L_{b2}\left(\frac{M}{M_{b2}} \right)^{\gamma_2} \left[\frac{1}{2}+ \frac{1}{2} \left(\frac{M}{M_{b2}} \right)^{1/\Delta_2} \right]^{\Delta_2(-\gamma_2+\gamma_3)} & M/M_{\odot} \geq 12 \\
\end{cases}
\end{equation}
\end{linenomath}
This relation was derived in \citet{menchiari} from merging existing parameterizations from \citet{yungelson2008} and \citet{eker2018}, and it can be applied to the MS evolution in a broad mass range (extending even beyond $100 M_\odot$). In Eq.~\eqref{eqn:Lstar} $L_{b1}=3191 \, L_\odot$, $L_{b2}=368874 \, L_\odot$, $M_{b1}=7 M_\odot$, $M_{b2}=36.089 M_\odot$, $\gamma_1=3.97$, $\gamma_2=2.86$, $\gamma_3=1.34$, $\Delta_1=0.01$, $\Delta_2=0.15$, and $\epsilon=0.817$. 
We further assumed that all member stars were located at the same distance as the cluster: Hence, the minimum magnitude member star per cluster allowed us to infer the maximum mass $M^*_{\rm max}$, while the mode magnitude determines the minimum stellar mass $M^*_{\rm min}$. The distributions of these values for the clusters we analyzed are also provided in Appendix~\ref{sec:appA}.

The number of member stars $N^*$ between the minimum inferred stellar mass $M^*_{\rm min}$ and the maximum inferred stellar mass $M^*_{\rm max}$ is known from observations and shown in Figure~\ref{fig:nstar}. The distribution of the number of stars per cluster has a median value of 58, which is in line with typical values of open clusters. Finally, the normalization $k$ of Equation~\eqref{eq:imf} was obtained by inverting the following expression:
\begin{linenomath}
\begin{equation}
  N^* = \int_{M^*_{\rm min}}^{M^*_{\rm max}} \xi(M) \, dM \, .
\label{eq:IMF_kN}
\end{equation}
\end{linenomath}

After normalizing the mass distribution of each cluster to match the Gaia number of observed stars, we obtained the total number of stars expected per cluster according to theoretical considerations, that is, we extended the following integral to compute the cluster mass in 
\begin{linenomath}
\begin{equation}
\label{eq:mc}
 M_{\rm c} = \int_{M_{\rm min}}^{M_{\rm max}} \xi(M) M \,dM \, 
\end{equation}
\end{linenomath}

\noindent
between a minimum $M_{\rm min}$ and maximum $M_{\rm max}$ stellar mass, in accordance with both stellar and cluster evolutionary theory. In particular, $M_{\rm min}$ was assumed to be an absolute minimum mass of $\sim0.08 \,M_\odot$, related to the formation of brown dwarfs, while $M_{\rm max}$ was determined by the following considerations. On one hand, the age of the cluster affects the highest value of the stellar mass that has not yet exploded into SN \citep{buzzoni2002}, under the hypothesis that the stellar formation process is instantaneous, such that the cluster age corresponds to the stellar age. In order to derive the relation of the maximum stellar mass allowed by its age, we adopted data from figure~3 of \citet{buzzoni2002}, collecting the turn-over time of a sample of MS stars with solar metallicity (representing the reference time for SN explosion) as a function of their mass. We parameterized and inverted the analytical relation describing the data of \citet{buzzoni2002}  according to the following expression:
\begin{linenomath}
\begin{equation}
\log_{10} [M_{\rm max}/M_\odot] = -C*\ln \left[\log_{10} [T_{\rm age}/{\rm yr}]/A-B \right] \, ,
\label{eq:param}
\end{equation}
\end{linenomath}
where the parameter values resulting from the fit are provided in Appendix~\ref{sec:appA}.

On the other hand, the maximum  mass of a member star is determined in the first instance by the parent cluster mass, namely since the time of its formation \cite[see Figure~1 in][]{weidner2004}, most likely as a result of the interplay between the stellar feedback and the binding energy of the cluster-forming molecular cloud core. 
The latter effect naturally embeds the highest stellar mass observed. However, in order to determine the maximum stellar mass expected per cluster with the \citet{Weidner-Kroupa:2006} approach, knowledge of the cluster mass is required, which (by definition) is unknown. We hence proceeded iteratively by first computing a zeroth-order estimate of the cluster mass $M_{\rm c,0}$, which we call seed cluster mass, which is obtained through Eq.~\eqref{eq:mc} taking $M_{\max}$ as the minimum between $150\,M_\odot$ and the highest stellar mass allowed by stellar evolution given the cluster age $T_{\rm age}$ (according to Eq.~\eqref{eq:param}). We hence proceeded by iterations aiming at refining the prediction about $M_{\max}$ by accounting for the relation with the parental cluster mass, as derived by \citet{weidner2004}.

\begin{figure}
\centering
\includegraphics[width=0.48\textwidth]{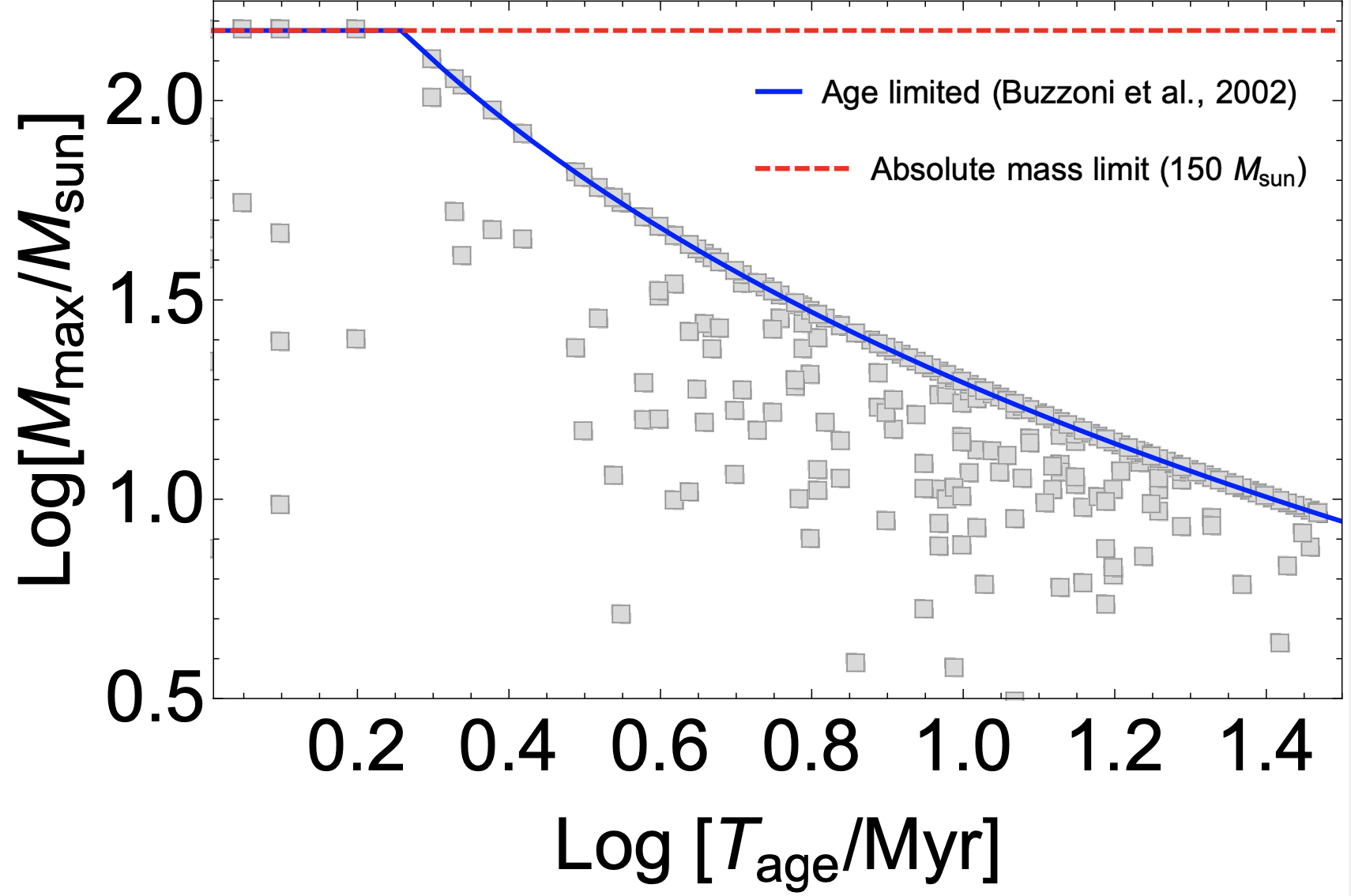}
\caption{Maximum expected stellar mass as a function of cluster age. The gray markers refer to the sample of selected young clusters from the Gaia catalog, whose maximum masses are evaluated from the iterative procedure described in the text. The solid blue line shows the result from Eq.~\eqref{eq:param}, and the dashed red line represents the highest stellar mass value observed in the Milky Way at $\simeq 150 M_\odot$.}
\label{fig:Mmax}
\end{figure}

\begin{figure}
\centering
\subfigure[\label{fig:MmaxStar}]{\includegraphics[width=0.47\textwidth]{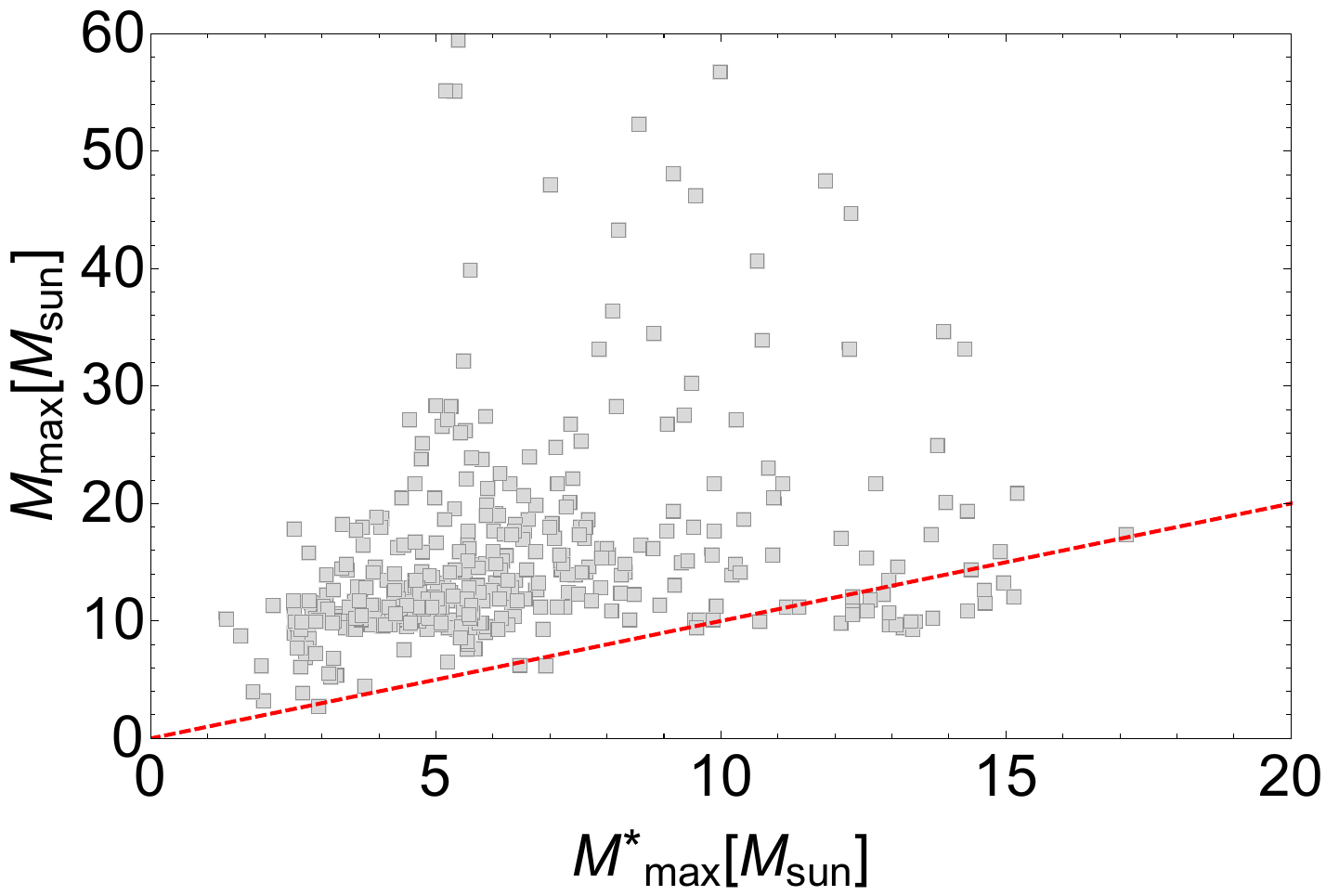}}
\subfigure[\label{fig:mass}]{\includegraphics[width=0.47\textwidth]{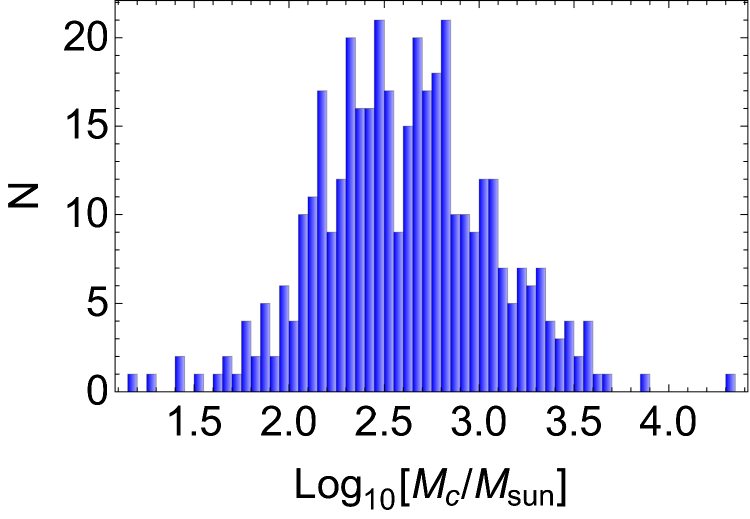}}
\caption{Estimated stellar and cluster masses from the Gaia DR2 young cluster sample. (a) Comparison of the maximum stellar mass per cluster obtained in our calculation (ordinate) and that inferred from the Gaia observation (abscissa). (b) Distribution of cluster masses in units of solar masses  for the selected young clusters.}
\label{fig:mass_mmax}
\end{figure}

Namely, for the clusters where $M_{\rm max}(T_{\rm age})>M_{\rm max}(M_{\rm c,0})$, we implemented an iterative approach based on the seed cluster mass values: We first derived the maximum stellar mass expected at the time of cluster formation according to \citet{Weidner-Kroupa:2006}, and then adopted this value to recompute an updated cluster mass through Eq.~\eqref{eq:mc}, repeating the calculation to obtain the corresponding $M_{\rm max}$ and so on, until after a few iterations, the procedure converged with an accuracy below 5\% with respect to the cluster masses evaluated in the previous step. The values of $M_{\max}$ obtained with this procedure are shown in Fig.~\ref{fig:Mmax} as filled markers, in comparison with the age-limited model from \citet{buzzoni2002}. 
In Fig.~\ref{fig:MmaxStar}, we compare these values to those we inferred from Gaia observations ($M_{\rm max}^*$). The latter are clearly always lower than what is expected from theory, except for a few cases in which the adopted cluster ages are doubtful, as we discuss further in the next section. 
We additionally compared the maximum stellar mass values determined for each cluster with the mass values of observed star members that are available from the StarHorse routine. We found systematically lower values in the latter case, as expected. 

The resulting cluster mass distribution is shown in Fig.~\ref{fig:mass}, with a median value of $\sim 413 \rm M_\odot$ and a standard deviation of $\sim 1472 \rm M_\odot$. The highest value estimated in the sample is $M_c^{\rm max}\simeq 2.2 \times 10^4 \, \rm M_\odot$ and corresponds to Westerlund 1. We note that the distribution of the logarithm of young cluster masses has a bell-like shape, which is expected to be induced by the completeness limit of the survey. The distribution can be fit with a Gaussian function with a mean value of $\log_{10} [M_\mu /M_\odot] = 2.6$ and a  standard deviation of $\sigma_{\log [M_\mu/ M_\odot]} = 0.45$. In the next section, we compare our mass estimates with other results in the literature, while in Sec.~\ref{sec:wr}, we describe how the mass uncertainty of our method was quantified based on the observed number of WR stars per cluster.

\section{Comparison with existing cluster mass calculations}
\label{sec:comparison}
In this section, we compare our estimate of cluster masses with recent results from other works. In particular, \citet{just2023} explored the latest MWSC data release that provided the so-called tidal radius $r_{\rm t}$, a spatial parameter first introduced by \citet{king1962} to empirically describe the radial profile of the projected surface density of a star cluster, specifically the point where the projected density drops to zero. The set of three King's parameters was later found to fully describe the density profile of a spherical system in a quasi-equilibrium configuration. The tidal radius is therefore commonly adopted to compute the mass of virialized systems, such as the long-lived globular clusters. The application to open clusters is in turn more uncertain because their dynamics is dominated by the the tidal forces of the Galaxy rather than by their internal potential. Nonetheless, \citet{just2023} have computed tidal masses of 2227 clusters, regardless of their nature. Of these, 505 are younger than 30~Myr, but only 149 star clusters are in common with the Gaia sample we investigated. By comparing the masses of  the common clusters, we found that the tidal approach yields masses that are systematically lower than those based on star counts used here. This behavior might be due to mass segregation effects \citep{PortegiesZwart:2010, Kirk+2011}, inducing smaller estimates of the cluster radius that might strongly bias the results toward lower masses \citep{ernst2010}. 

A more proper comparison can be performed with a recent analysis by \citet{almeida2023} that included a cluster mass evaluation for 773 open clusters from Gaia DR2. The authors developed a novel strategy based on the simulation of a synthetic cluster population followed by an isochrone fitting procedure that enabled them to constrain cluster ages, distances, masses, and binary fractions. This time, the common cluster sample (younger than 30 Myr) contains 102 objects, for which we investigated the differences not only in mass, but also in the other parameters derived by the authors. With respect to the parameters provided by the Gaia/MWSC catalogs, the distances found by \citet{almeida2023} agree well for cluster distances smaller than 2~kpc. In contrast, for farther clusters \citet{almeida2023} obtained significantly smaller distances than Gaia/MWSC catalogs. More interestingly, the revised age estimates largely disagree with the values provided by Gaia/MWSC given that the age distribution in the \citet{almeida2023} sample is quite narrow and peaked around $(7.2 \pm 0.30)$~Myr. These results indicate that the age values provided by the Gaia/MWS catalogs might be underestimated in the case of younger clusters (<6~Myr), but overestimated for the older ones. This behavior is also supported by the investigation of the observed versus expected WR stars in clusters that we present in Sec.~\ref{sec:wr}. 

\begin{figure}
\centering
\includegraphics[width=\columnwidth]{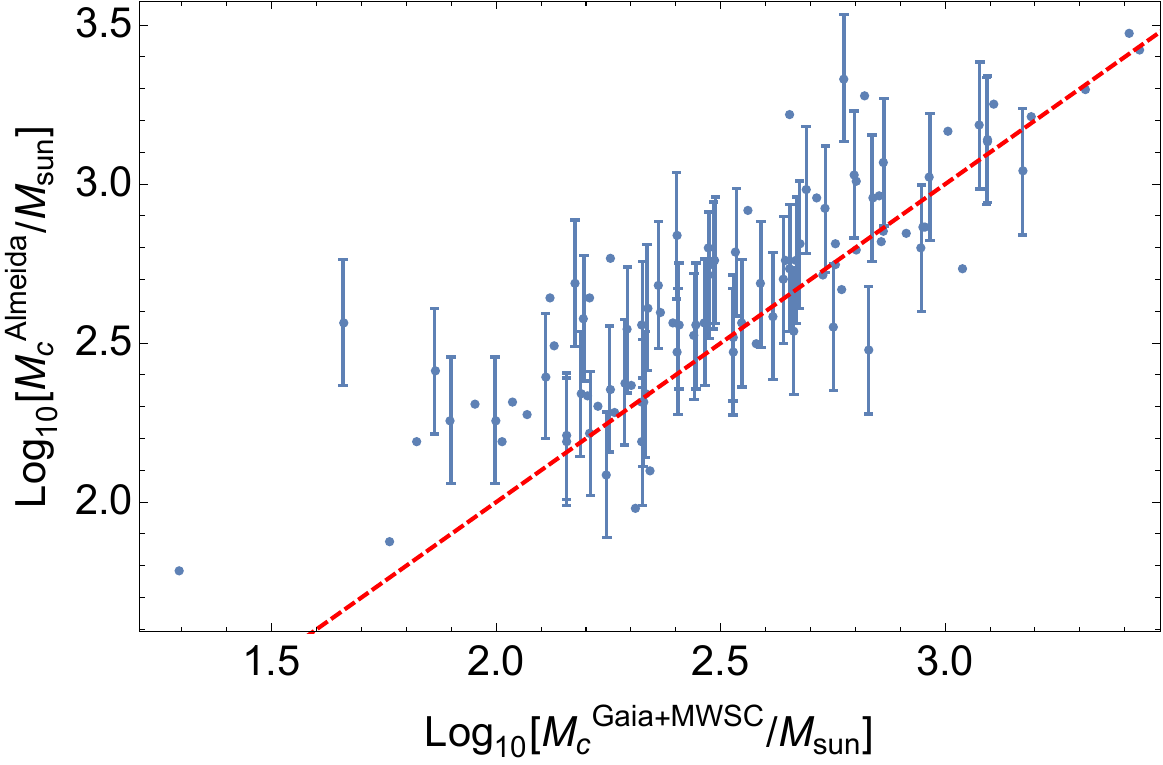}
\caption{Comparison of the cluster masses estimated in this work (abscissa) and by \citet{almeida2023} (ordinate) for the common sample of 102 young clusters from Gaia DR2. The dashed red line shows the identity.}
\label{fig:almeida}
\end{figure}

The masses of single clusters agree overall within the uncertainties, but the average mass estimated by \citet{almeida2023} is higher by $\sim 40\%$ than ours, as shown in Figure~\ref{fig:almeida}. This is expected because the two methods often rely on different cluster physical parameters to determine their masses. In particular, the method developed by \citet{almeida2023} also accounts for binary stars, which we neglected. However, our sample is larger than that used by \citet{almeida2023}, and even more importantly, it includes stellar clusters with higher masses. This is crucial for the computation of gamma-ray emission from these objects, providing us with a more realistic estimate of their role as particle accelerators at the highest energies.

Finally, in Table~\ref{tab:SCmass} we compare our results for a few of the most massive star clusters of our sample with the mass and age values found in other works. The cited works usually estimated the cluster masses by evaluating their stellar content with a joint analysis of optical and near-infrared imaging. Our estimates are in general lower (by a factor $\sim 2$ in three out of seven cases and a factor $\sim 15$ in one case), with the exception of one case (Danks~2). We show in the next section that the number of WR stars also suggests that we underestimate the masses of the most massive stellar clusters by a factor $\sim 3$ at least, as demonstrated below. 

\begin{table}[tp]
\hspace{-0.2cm}
    \begin{tabular}{l|cc|ccl}
    Cluster & $M_{\rm c}$ & $T_{\rm age}$ & $M_{\rm c}$ & $T_{\rm age}$ & Ref. \\
                  & [$10^3$M$_{\odot}$] \hspace{-0.3cm} & [Myr] & [$10^3$M$_{\odot}$] \hspace{-0.3cm} & [Myr] &  \\
    \hline\hline
    Wd 1         & 22.2 & 7.9  & $49$ & $4.0\pm 0.5$ & [1]  \\
    Danks 2      & 7.8  & 1.0  & $3.0 \pm 0.8$  & $3.0^{+3.0}_{-1.0}$ & [2]\\ 
    NGC 3603 \hspace{-0.1cm}    & 4.8  & 1.0  & $13\pm 3$ & 1-2 & [3]\\
    Danks 1      & 4.4  & 1.0  & $8.0 \pm 1.5$ & $1.5^{+1.5}_{-0.5}$ & [2]\\
    NGC 6231 \hspace{-0.1cm}    & 3.2  & 13.8 & $3.75\pm0.45$   & 2--7 & [4]\\
    Wd 2         & 2.2  & 4.0  & 3.6  & 2.5 & [5]\\   
   NGC 6611 & 1.7 & 2.14 & 25 & 2-5 & [5]\\   
    \hline    
    \end{tabular}
    \caption{Comparison of our estimates of mass (second column) and the values provided by other works (fourth column) for a subset of clusters. The cluster ages are also reported, namely the values we considered from Gaia/MWSC catalog (third column) and the values usually adopted in the literature (fifth column). The references are indicated in the last column and correspond to [1]= \citet{gennaro2011}; [2]= \citet{Davies+2012-Danks12}; [3]= \citet{Harayama+2008-NGC3603}; [4]= \citet{Kuhn+2017-NGC6231}; [5]= \citet{Pfalzner:2009-Wd2}.}
    \label{tab:SCmass}
\end{table}

\section{Estimating the cluster mass uncertainty using WR stars}
\label{sec:wr}
The cluster masses estimated in Sec.~\ref{sec:stars} should be taken as a lower limit for two reasons. First, we neglected the stellar component in binary systems, which may cause us to underestimate the cluster mass by a few tens of percent. For instance, the recent estimate by \cite{Hunt-Reffert:2023} \citep[see also][]{Hunt_PhD:2023} suggested that the fraction of unresolved binaries can increase the cluster mass by 10\%-30\%. Second, the identification of cluster members by \cite{gaia2020} was achieved through a specific neural network algorithm. However, techniques based on machine learning strongly depend on the initial set used to train the neural network. Other works that were based on different neural network architectures and had different training sets identified more stars \cite[even up to a factor of 2; see, e.g.,][]{mem2021,van2023}, which clearly would impact any mass estimation method based on the star count. Moreover, \cite{Buckner+2023} applied single star identification criteria to mock clusters, showing that for distances $\gtrsim 1$\,kpc, they missed up a fraction $>50\%$ of the cluster members. Therefore, it would be highly desirable to have an independent estimate of the cluster masses. This can be obtained by comparing the number of WR stars associated with star clusters with those expected from theory, as explained below.

\begin{figure}
\centering
\includegraphics[width=\columnwidth]{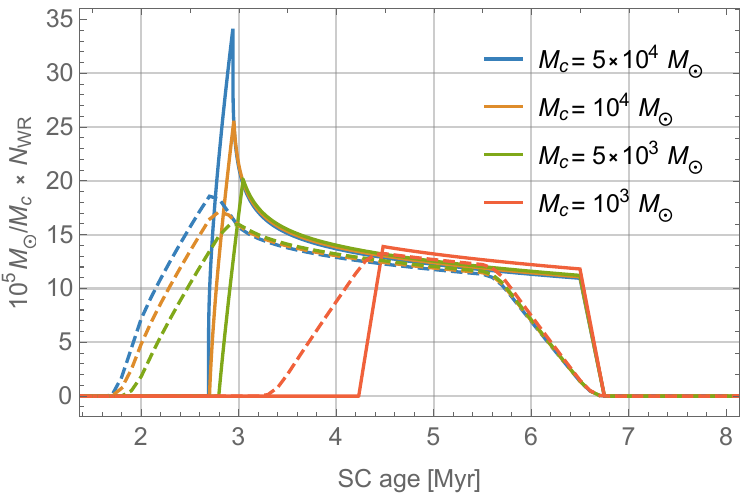}
\caption{Theoretical prediction for the number of WR stars, $N_{\rm WR}$, inside a cluster as a function of its age and for different cluster masses as labeled. The duration of the SF process is assumed to be 0 and 1 Myr for the solid and dashed lines, respectively. $N_{\rm WR}$ is divided by $M_{\rm c}/(10^5\,\rm M_{\odot})$.}
\label{fig:WRtheory}
\end{figure}

To provide a theoretical estimate for the number of WR stars, we assumed that at the end of the MS phase, all stars with a mass above $M_{\rm WR, \min}$ experience the WR phase. This minimum mass was estimated from stellar evolution codes to be between 22 and 37 M$_\odot$ \citep{Eldridge-Vink:2006}. The average duration of the WR phase was estimated from stellar evolution models to be $\approx [0.1-1]$\,Myr, depending on several parameters such as the initial mass, metallicity, and stellar rotation \citep{Meynet+1994, Meynet-Maeder:2005}. Because of the large theoretical uncertainties, we adopted the range inferred from the observed WR stars, that is, $\Delta t_{\rm WR} = [0.25\--0.40]$\,Myr \citep{rosslowe2015}. Hence, under the assumption that the star formation process in the cluster is instantaneous and that the duration of the MS phase is regulated by Eq.~\eqref{eq:param}, the number of WR stars is given by all stars whose MS phase lasted between $t-\Delta t_{\rm WR}$  and $t$ ($t$ being the cluster age). This translates into a mass interval between $M_{\rm WR,\min}(t)= \min[ M_{\rm WR, \min}, M(t_{\rm MS}=t)]$ and $M_{\rm WR,\max}(t)= M(t_{\rm MS}=t - \Delta t_{\rm WR})$.
In reality, the star formation (SF) process is not instantaneous, but lasts for a time $\Delta T_{\rm SF}$ that mainly depends on the structure of the parent molecular cloud from which the cluster is born. Numerical simulations showed that this time can last up to $\sim 3$\,Myr \citep{Krause+2020SSRv} and that the star formation rate is not constant during this time. Here, we accounted for the delayed SF process, while assuming that the star formation rate (SFR) remained constant during the entire period. Under these assumptions, the total number of WR stars as a function of cluster mass and age is
\begin{linenomath}
\begin{equation}
\label{eq:WRnumber}
    N_{\rm WR}(M_{\rm c},t) = \frac{1}{\Delta T_{\rm SF}} \int_{0}^{\Delta T_{\rm SF}} dt' \int_{M_{\rm WR,\min}(t-t')}^{M_{\rm WR,\max}(t-t')} \xi(M_{\rm c}, M) \,dM \, ,
\end{equation}
\end{linenomath}
where we explicitly introduce the dependence of $\xi$ on the total cluster mass $M_{\rm c}$. Fig.~\ref{fig:WRtheory} shows the result of solving Eq.~\eqref{eq:WRnumber} for different cluster masses, fixing $M_{\rm WR, \min} = 25 M_{\odot}$ and using two different values for $\Delta T_{\rm SF}$, namely 0 (instantaneous star formation), and 1 Myr. In the former case, $N_{\rm WR}(t)$ shows a peak close to the threshold age when the most massive star in the cluster enters the WR phase, while in the latter case, this peak is smoothed out. The cutoff at $t\simeq 6.5$\,Myr corresponds to the MS duration of stars with a mass of $25\,\rm M_{\odot}$.

At the time of writing, the Galactic WR catalog \citep{rosslowe2015}\footnote{See the online catalog at \url{http://pacrowther.staff.shef.ac.uk/WRcat/}, last update from June 2023.} counts 669 WRs, 49 of which are assumed to be associated with 14 of the young clusters in our sample, namely Ruprecht~44, Westerlund~1 and 2 (Wd~1 and Wd~2), Trumpler~16, Sher~1, NGC~3603, Hogg~15, Danks~1 and 2, NGC~6231, Dolidze~3, Markarian~50, and Berkley~86 and 87. In descending order, Wd~1 is associated with 24 WRs (second only to the Galactic center), followed by Danks~1 with 6 WR stars, NGC~3603 with 5 WRs, and Wd~2 with 2 WRs. The remaining clusters are associated with a single WR. 

We compared the predicted number of WRs by applying Eq.~\eqref{eq:WRnumber} to the cluster sample using the lower limits of the masses estimated in Sec.~\ref{sec:stars} and the associated ages. This procedure resulted in a predicted number of WRs equal to zero, regardless of the value of $\Delta T_{\rm SF}$. This result may be due to both an underestimated value of the cluster masses and/or an incorrect estimate of their ages. The latter, in particular, is supported by several pieces of evidence, including Fig.~\ref{fig:WRtheory} and the discussion in the previous section. As shown in Fig.~\ref{fig:WRtheory}, the clusters would be expected to host WR stars in the age range between $\sim 2.5$ and $6.5$~Myr, depending on their total mass. By assuming the catalog ages, we should not observe WRs in some cases in which they are reported. The only solution to this problem is a revision of their age estimate.
However, an incorrect age estimate cannot entirely justify the discrepancy. Even when we assume that all clusters have an age of $\sim 4$\,Myr (to maximize the number of WR stars), the theoretical number of WRs expected in our cluster sample would be 11 (3 of which lie in Wd~1) compared to 49 (24 of which lie in Wd~1).  We interpret this result as further evidence that the cluster masses evaluated in Sec.~\ref{sec:stars} are underestimated.

While confirming that the total cluster masses are underestimated, the observed number of WR stars provides us with a method for quantifying the mass uncertainty. By assuming that the actual cluster masses are a factor $\chi$ higher than the value estimated in Sec.~\ref{sec:stars}, we aim at reproducing the observed WR number. 
However, before proceeding with a quantitative estimate of $\chi$, we note that neither the star cluster catalog nor the WR catalog are complete. We evaluated the completeness of the former in Sec.~\ref{sec:data} to be about 2-3~kpc. The WR catalog can in turn be considered complete up to magnitude 8 in $\rm K_s$ band \cite[see \S~ 4.3.1 and references therein]{rosslowe2015}. This roughly corresponds to a distance of 3~kpc in the Galactic plane \cite[see, e.g.,][]{Kanarek+2015}. To be conservative, we restricted our analysis to WR/cluster associations closer than 2.5 kpc. Within this distance range there are five associations with a total of seven WRs.

\begin{figure}
\centering
\includegraphics[width=\columnwidth]{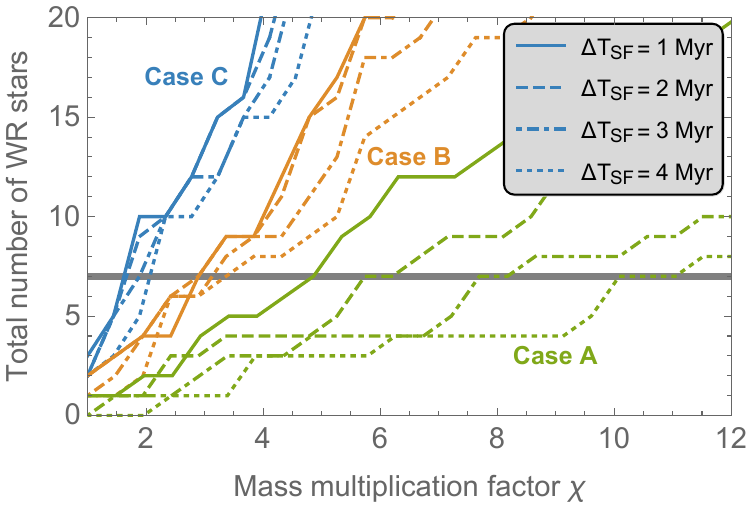}
\caption{Comparison of the observed number of WR stars in our cluster sample and the theoretical predictions where the cluster mass is multiplied by an arbitrary factor. The different line styles refer to different values of the duration of the star formation process, as indicated in the legend, and for three different cases: A (green), B (orange), and C (blue). The horizontal line corresponds to the number of observed WRs in the cluster sample, restricted to systems closer than 2.5~kpc, where both the Gaia and WR catalogs are assumed to be complete. See text for details.}
\label{fig:WRcomparison}
\end{figure}

Fig.~\ref{fig:WRcomparison} reports the predicted number of WR stars in clusters within 2.5~kpc as a function of $\chi$ and for different values of $\Delta T_{\rm SF}$. 
To evaluate the impact of different assumptions, we show three sets of curves corresponding to different assumptions,
\begin{itemize}
    \item {\it Case A}: $\Delta t_{\rm WR} = 0.25$\,Myr and $M_{\rm WR,\min} =  37 \,\rm M_{\odot}$; 
    \item {\it Case B}: $\Delta t_{\rm WR} = 0.40$\,Myr and $M_{\rm WR,\min} = 22 \,\rm M_{\odot}$;
    \item {\it Case C}: as case B, but with all clusters with estimated ages younger than 10\,Myr set to be equal to 4\,Myr.
\end{itemize}
Case C was included to account for the uncertainty in the age estimate for very young clusters to maximize the number of WRs with respect to the cluster age. It should therefore be considered an extreme case. The result for case A suggests that our method underestimates the cluster masses by a factor $\gtrsim 5$, while for case B, we have $\chi \simeq 2.8-3.3$, depending on the value of $\Delta T_{\rm SF}$. The most optimistic of all cases, case C, requires $\chi \simeq 1.6-2.0$. In addition to the total number of WR stars, we also compared the mocked distribution of clusters with at least one WR with the observed one by performing a Kolmogorov-Smirnov test. The estimated $p$-value is $\lesssim 0.4$ for both cases A and C, and it is $\gtrsim 0.6$ for case B, suggesting that the latter is more compatible with the observed distribution than cases A or C.
Finally, we note that when we neglected the completeness issue and applied the same analysis to the whole star cluster and WR samples, we obtained similar conclusions to those shown in Fig.~\ref{fig:WRcomparison}.

A few additional comments are in order. The analysis discussed above may be affected by several additional uncertainties. From an observational point of view, WR stars might erroneously be associated with star clusters, as discussed, for example, in \citet{rate2020}. However, it seems unlikely that these misassociations apply to the majority of clusters, especially to those closer than 2.5 kpc where the parallax estimate is reliable.
On the other hand, the theoretical approach we developed is quite simplified, especially concerning the physics of WR stars. The stars that undergo a WR phase and how long this phase lasts remain highly debated questions. However, the assumed intervals for $M_{\rm WR,\min}$ and $\Delta T_{\rm WR}$ cover the ranges usually adopted in the literature. A further increase in the number of the WR stars would require either $M_{\rm WR,\min} < 22\, \rm M_{\odot}$ or $\Delta T_{\rm WR}> 0.4$\,Myr.  
Another possibility to obtain a larger number of WRs is to change the IMF into a top-heavy distribution for high masses (i.e., harder than the Salpeter scaling $M^{-2.3}$ assumed here), which would result in a higher number of massive stars. This may be the case for very massive star clusters such as those observed in the Galactic center, for instance, Arches \citep{Hosek+2019} and even Wd1 \citep{Lim+2013-Wd1}, but not in general because the IMF of local clusters agrees well with a Salpeter distribution \citep{Bastian+2010}.
Most likely, the less well-constrained parameter that also affects our mass estimate is the value of cluster ages, which is difficult to obtain by isochrone fitting alone, particularly for young clusters. As we showed in Sec.~\ref{sec:comparison}, the age values available in the literature differ significantly from the values quoted by \citet{gaia2020} for at least some massive clusters. Nonetheless, even when this uncertainty was completely removed (as for case C), the cluster masses would still need to be increased by at least 60\%.

Overall, we estimate an uncertainty for the cluster mass lower limits derived above of at least a factor {$\sim 3$} towards higher values.

\section{A lower limit to cluster wind luminosities}
\label{sec:lum}
Based on the above considerations, we proceeded to compute the cluster mass-loss rate by summing the mass-loss rates of all its member stars $\dot{M}_{\rm s}(M)$ as
\begin{linenomath}
\begin{equation}
\label{eq:mdotc}
 \dot{M}_{\rm c} = \int_{M_{\rm min}}^{M_{\rm max}} \xi(M) \dot{M}_{\rm s}(M) \,dM \, .
\end{equation}
\end{linenomath}
For the stellar mass-loss rate, we relied on the analytical and empirical prescription given by \cite{nieu1990} relative to MS stars, depending on stellar luminosity $L_s(M)$ and radius $R_s(M)$,
\begin{linenomath}
\begin{equation}
\label{eqn:mloss}
  \dot{M}_{\rm s}(M) \simeq 9.55 \times 10^{-15} \left(\frac{L_{\rm s}(M)}{L_\odot} \right)^{1.24} \left(\frac{M}{M_\odot} \right)^{0.16} \left( \frac{R_{\rm s}(M)}{R_\odot} \right)^{0.81} \, \frac{\rm M_\odot}{{\rm yr}},
\end{equation}
\end{linenomath}
while the dependence of the stellar radius on mass was taken from \citet{Demircan1991}, resulting in
\begin{linenomath}
\begin{equation}
\label{eqn:rstar}
R_{\rm s}(M)=0.85 \, R_\odot \, (M/M_\odot)^{0.67}.
\end{equation}
\end{linenomath}
The prescription adopted for the stellar mass-loss rate in Eq.~\eqref{eqn:mloss} was chosen because its empirically driven nature fits the purposes of a statistical study, however it does not account for the possible clumpiness of stellar winds. Clumps can enhance the strength of the emission lines, which depends on the density squared (e.g., H$_{\alpha}$ lines), implying a lower value of the inferred mass-loss rate \citep{renzo}. However, if the clumps are dense enough, they could become optically thick to the emitted radiation, suppressing the emission \cite[see discussion in][]{Vink_rew:2022}. The dominant effect remains unclear so far, and more investigation of the topic is required. For instance, in a combined analysis of the H$_{\alpha}$ and P v doublet lines detected from $\zeta$ Puppis, \citet{Oskinova+2007} found that the two effects compensate almost perfectly and that the inferred mass-loss rate is very close to the rate predicted without accounting for clumpiness. Hence, in the following discussion, we simply rely on the empirical prescription for the mass-loss rate given in Eq.~\eqref{eqn:mloss} and do not account for clumpiness.

The other important parameter is the wind speed, $v_{\rm w,s} (M)$, whose value is related to the escape velocity from the star, 
\begin{linenomath}
\begin{equation}
v_{\rm esc}(M)= \sqrt{\frac{2 G M}{R_{\rm s}(M)}} \, ,
\end{equation}
\end{linenomath}
$G$ being the gravitational constant. With respect to $v_{\rm esc}$, the wind speed includes two additional effects. One is the presence of metals, which increases the effect of photon pressure (line-driven winds), and the second effect depends on the density of the wind itself, which acts by reducing the radiation pressure. These effects are taken into account by introducing a multiplication factor $C$ and the Eddington luminosity of the star $L_{\rm Edd} (M) =4\pi G m_{\rm p} M c/\sigma_{\rm T}$, with the Thomson cross section $\sigma_{\rm T}$ and proton mass $m_{\rm p}$, such that the stellar wind speed is given by \citep{puls2000}
\begin{linenomath}
\begin{equation}
\label{eqn:vws}
v_{\rm w,s}(M) = C(T_{\rm eff}) \sqrt{ \frac{2 G_{\rm N} M}{R_{\rm s}(M)} \left(1-\frac{L_{\rm s}(M)}{L_{\rm Edd}(M)} \right) } \,.
\end{equation}
\end{linenomath}
The coefficient $C$ depends on the physical mechanism determining the wind pressure and is a function of the stellar effective temperature $T_{\rm eff}$,
\begin{linenomath}
\begin{equation}
C(T_{\rm eff}) = 
\begin{cases}
1.0 & T_{\rm eff}/{\rm K}<10^4 \\
1.4 & 10^4 \leq T_{\rm eff}/{\rm K}<2.1 \times 10^4\\
2.65 & T_{\rm eff}/{\rm K} \geq 2.1 \times 10^4 \, .
\end{cases}
\end{equation}
\end{linenomath}
To avoid sharp discontinuities, we adopted a linear interpolation between C=1 an C= 2.65. The latter value represents line-driven stellar winds, for example, where metal deexcitation contributes to the wind pressure such that the wind speed is expected to be higher than the escape speed at infinity. 

Finally, it is possible to derive the cluster wind luminosity $L_{\rm w,c}$ through momentum conservation,
\begin{linenomath}
\begin{equation}
\label{eq:pc}
\dot{M}_{\rm c} v_{\rm w,c} = \int_{M_{\rm min}}^{M_{\rm max}} \xi(M) \dot{M}_{\rm s}(M) v_{\rm w,s}(M) \, dM \, ,
\end{equation}
\end{linenomath}
which, once inserted in the expression for the cluster wind luminosity, gives
\begin{linenomath}
\begin{equation}
\label{eq:lwc}
 L_{\rm c,w} = \frac{1}{2} \dot{M}_{\rm c} v^2_{\rm w,c} = \frac{1}{2} \frac{ \left(\int_{M_{\rm min}}^{M_{\rm max}} \xi(M) \dot{M}_{\rm s}(M) v_{\rm w,s} (M) dM \right)^2}{\int_{M_{\rm min}}^{M_{\rm max}} \xi(M) \dot{M}_{\rm s}(M) dM} \, .
\end{equation}
\end{linenomath}

\begin{figure}
\centering
\subfigure[\label{fig:mdot}]{\includegraphics[width=0.47\textwidth]{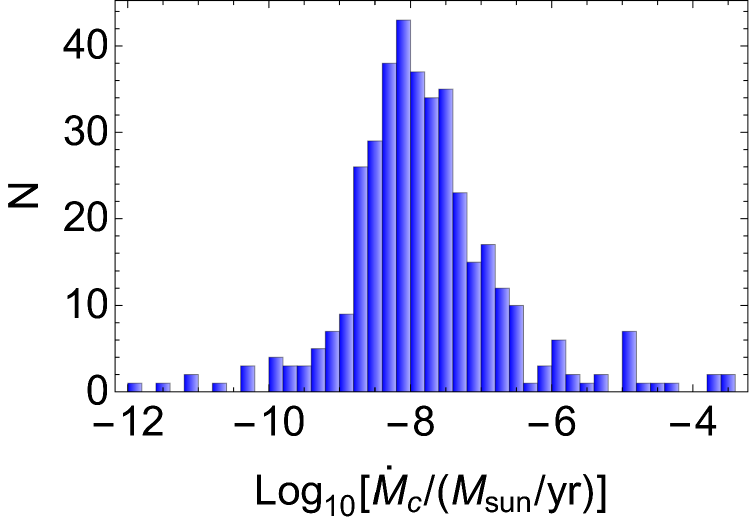}}
\subfigure[\label{fig:lum}]{\includegraphics[width=0.46\textwidth]{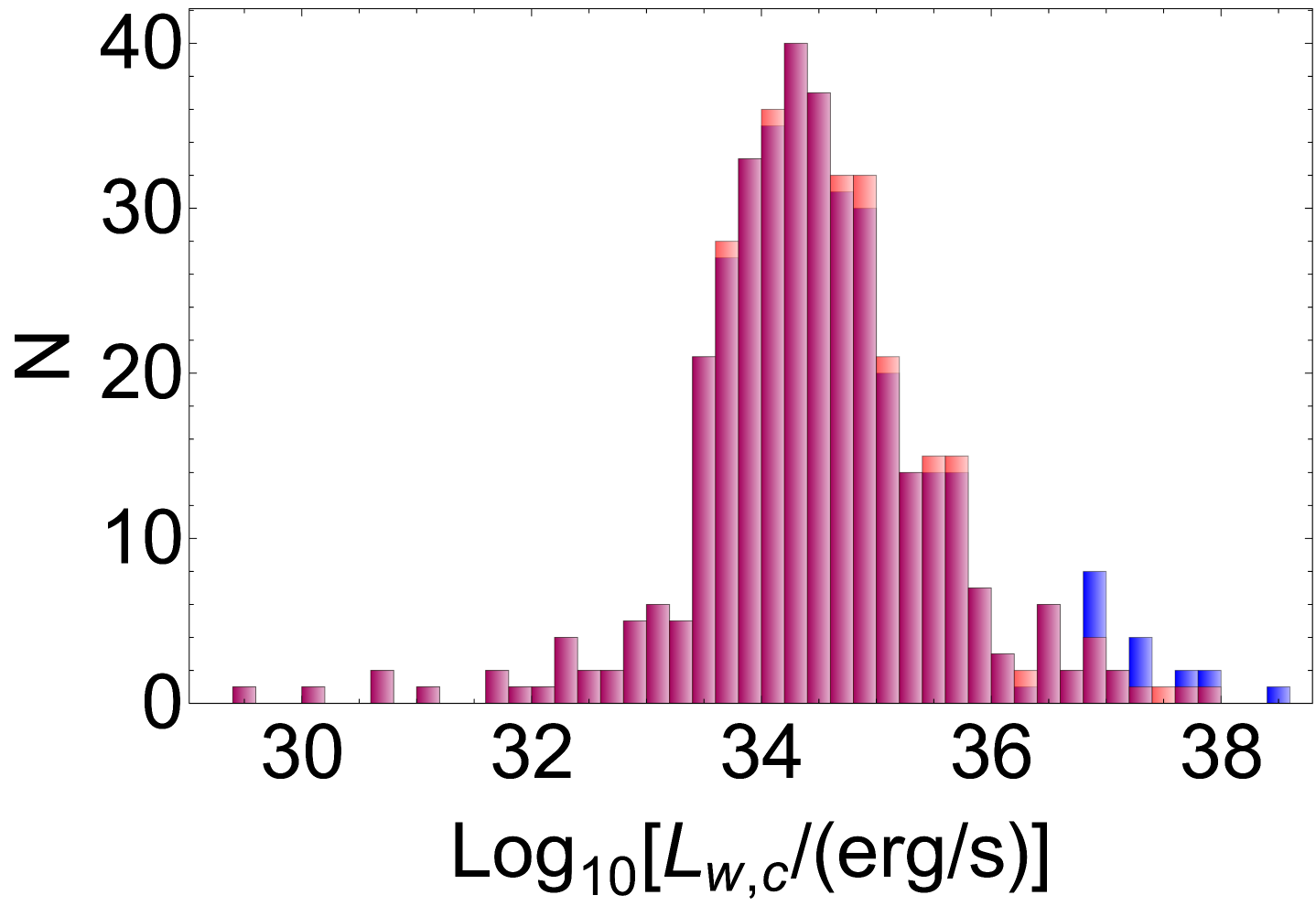}}
\caption{Distribution of the cluster mass-loss rates (panel a) in units of solar masses per year and of the wind luminosity (panel b) in units of erg/s for the selected young clusters. The latter shows the cluster wind luminosity computed both with (blue) and without (red) the contribution of observed WRs. The shaded part of the histogram represents overlapping values.}
\label{fig:lum_hist}
\end{figure}

In addition to MS stars, we included the contribution of the observed WR stars to the mass-loss rate and wind luminosities of clusters accounting for the 49 observed WR stars associated with the clusters of our sample. Since the catalog also provides information concerning the spectral class of each star (WN, WC, WO, and their subclasses), we defined the values of the mass-loss rate and speed of their winds. Following \citet{crowther2007}, we assumed an average mass-loss rate of $10^{-4.9} \, M_\odot$/yr, regardless of the spectral type, and wind speeds amounting to $1.6\times 10^8$~cm/s and $2.3\times 10^8$~cm/s for WN and WC/WO types, respectively. We assigned these values to the observed WRs associated with our cluster sample, obtaining wind luminosities of $9.3\times 10^{36}$~erg/s and $2.3\times 10^{37}$~erg/s for WN and WC/WO types, respectively, which we include in the luminosity computation of the clusters. As a result, we found that the most luminous open cluster of our sample is Westerlund~1. 

The resulting cluster mass-loss rate and wind luminosity distributions are shown in Figs.~\ref{fig:mdot} and \ref{fig:lum}. The median values are $1.3 \times 10^{-8} \, \rm M_\odot \, yr^{-1}$ and $3 \times 10^{34}\, \rm erg\,s^{-1}$, but the high-energy tail extends up to $\sim 3 \times 10^{38}\, \rm erg\,s^{-1}$. For clarity, the panel with the wind luminosity shows histograms obtained both with and without the contribution of the observed WRs, indicating the crucial role of this stellar population in achieving higher energetics of the system. We additionally note that this result should be taken with the caveat that the star cluster and the WR samples are not complete at the same level. We may therefore have missed some WR stars associated with clusters in principle, especially for those located at distances $\gtrsim 3$\,kpc.
Based on the discussion about the cluster masses, we stress that the result for the wind luminosity should also be regarded as a lower limit.  

As a final comment, we note that the above expressions for the wind speed and mass-loss rate assume solar metallicity. Young stellar clusters, however, presumably have a higher metal content than the Sun. It might therefore be wondered how the wind luminosity is affected. A proper description of the metallicity dependence goes beyond the aim of the present work because it is a nontrivial task to determine this. The metallicity has only been measured for a few star clusters, and this measure often only relied on one star or a few stars. We limit ourselves to comment on its possible impact. Both $\dot{M}$ and $v_w$ depend on metallicity. \citet{Vink-Sander:2021} found that massive O stars have a mass-loss rate $\propto Z^{0.42}$, while the wind speed is $v_{\rm w} \propto Z^{0.19}$, implying that the final wind luminosity is $L_{\rm w} \propto Z^{0.8}$. For the less massive B-type stars, the dependence is far weaker. Considering that O-type stars dominate the wind power, we can state that if the average metallicity of YMSCs were $\sim 50\%$ higher than in the Sun, the total luminosity would increase by $\sim 40\%$.

The full compilation of cluster masses, mass-loss rates, and wind luminosities is reported in Table~\ref{tab:cluster_parameters} for further use. 

\section{Conclusions}
\label{sec:concl}
Very high-energy and ultra high-energy gamma-ray observations of stellar clusters to date only enumerate a few bright systems in our Galaxy, which include Wd~1 and 2, the Cygnus Cocoon, and W43 \citep{felixCluster,lhaaso2023}. The nature of their energetic emission is still unclear, even though the hadronic scenario is preferred \citep{felixCluster,emmaWd2}. In a few cases, the reported emission requires protons and nuclei to be accelerated up to $\sim$PeV energies. 

From a theoretical perspective, assessing the role of star clusters as extreme particle accelerators at a population level relies on the estimation of their masses and wind luminosities because they affect the energetics of non-thermal particles. Nonetheless, obtaining mass estimates of open clusters is a very difficult task, particularly for the young systems we are interested in, to probe the effectiveness of the WTS scenario. Young clusters are indeed more uncertain than older clusters, both in the use of stellar population synthesis models and observationally because velocity dispersion measurements are frequently lacking. Hence, finding the most reliable technique for estimating their mass remains an open issue in the astronomical community that is very much debated and often dependent on the system parameters. 

We have developed a mass estimation method for star clusters based on the properties of their individual member stars. We applied it to the Gaia DR2 open cluster catalog and obtained cluster masses ranging from $\sim10$ to $\sim10^4$ solar masses, in agreement with the expectations for the disk population of the Milky Way. The highest mass value obtained for the systems in our sample is for Wd~1, with a lower limit of $M_{\rm Wd1} \geq 2.2\times 10^4\, \rm M_\odot$. Literature studies with a completeness-corrected IMF reported its mass to be $4.91^{+1.79}_{-1.49} \times 10^4 \, M_\odot$ \citep{gennaro2011}, which is higher by a factor $\sim 2$ than our estimate. It is worth highlighting that strong evidence of mass segregation was found in this cluster, despite its young age of $4.0 \pm 0.5$~Myr \citep{gennaro2011}. 

We have further explored the usage of WR stars to quantify the uncertainty on the estimated masses. By developing a simplified model for WR occurrence in the star clusters of our Galaxy and comparing it to observations, we conclude that at least a factor 3 toward higher values should be considered for a realistic estimate of the cluster masses. This factor encloses several limitations of our procedure, which neglects binary systems as well as unresolved stars and cluster members that might belong to the cluster, but were not identified by the membership algorithms \citep{Buckner+2023}. This possibility is further supported by the comparison with alternative mass estimation methods available in the literature, which are restricted to a smaller cluster sample and lower cluster mass range, however. From both the WR estimation method and the comparison with cluster masses available in the literature, evidence is given that some of the cluster ages claimed in the catalogs are also unsuitable to describe their properties.

The estimated masses were used to also calculate the total wind luminosity, which is the most important parameter for establishing the gamma-ray luminosity for young stellar clusters, and which will be used in two companion papers to evaluate the detection prospects of known clusters in the gamma-ray band.

\begin{acknowledgements}
We thank G.~Sacco and R.~Crocker for useful discussions about cluster properties. SC gratefully acknowledges partial funding from the Bavarian State Ministry of Science and the Arts under the 2022 Visiting Professor Programme at the Erlangen Centre for Astroparticle Physics. AM is supported by the Deutsche Forschungsgemeinschaft (DFG, German Research Foundation) -- Project Number 452934793.
SM and GM are partially supported by the INAF Theory Grant 2022 {\it ‘‘Star Clusters As Cosmic Ray Factories''} and INAF Mini Grant 2023 {\it ‘‘Probing Young Massive Stellar Cluster as Cosmic Ray Factories''}.
\end{acknowledgements}

%
\bibliographystyle{aa} 
\bibliography{bibliography} 
%

\begin{appendix} 
\section{Normalization of the cluster stellar mass function}
\label{sec:appA}
\begin{figure}
\centering
\includegraphics[width=0.49\textwidth]{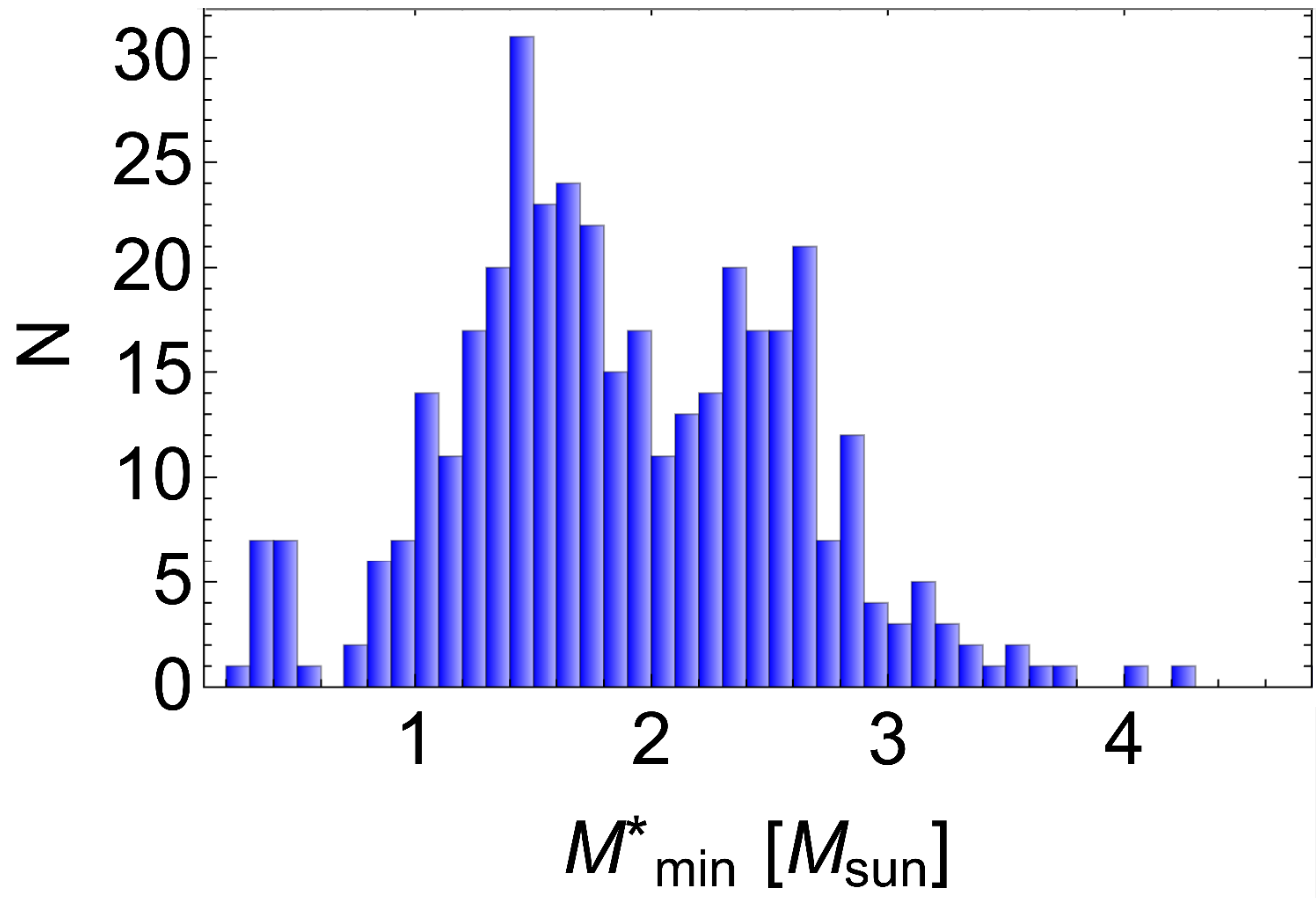}
\caption{Distribution of the minimum stellar masses as inferred from Gaia observations, described in Sec.~\ref{sec:stars}, for the clusters in the selected sample.}
\label{fig:mmin}
\end{figure}

\begin{figure}
\centering
\includegraphics[width=0.5\textwidth]{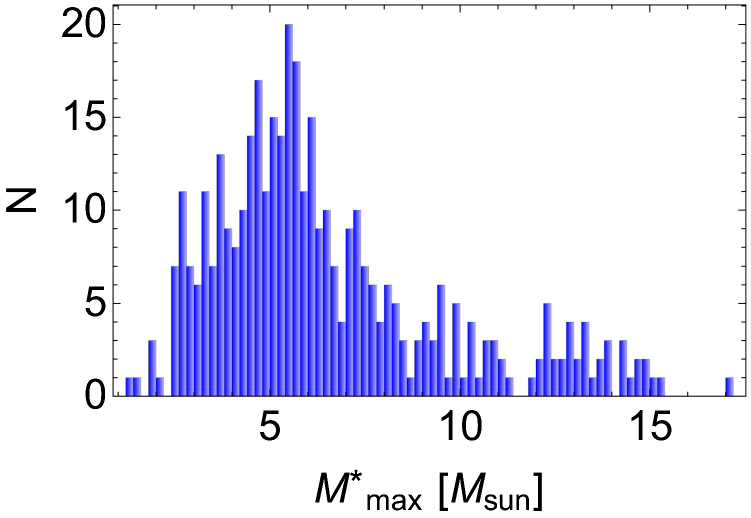}
\caption{Distribution of the maximum stellar masses as inferred from Gaia observations, described in Sec.~\ref{sec:stars}, for the clusters in the selected sample.}
\label{fig:mmax}
\end{figure}

\begin{figure}
\centering
\includegraphics[width=0.45\textwidth]{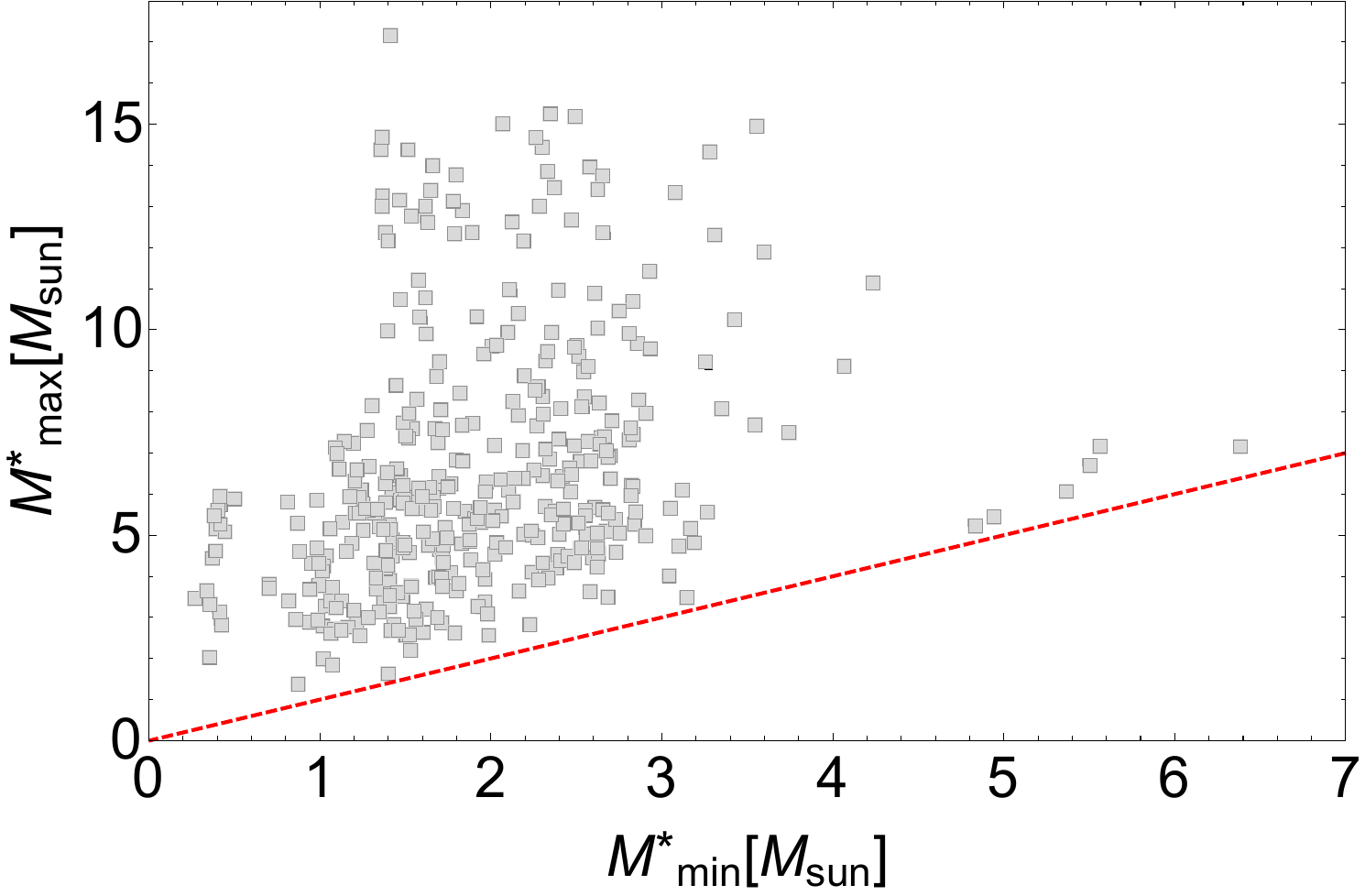}
\caption{Scatter plot between the minimum and maximum stellar mass values inferred from Gaia observations as described in Sec.~\ref{sec:stars} for the clusters in the selected sample. The dashed red line represents the bisector.}
\label{fig:smm}
\end{figure}

As discussed throughout the text, the method developed here for estimating cluster masses relies on the assumption of a stellar mass distribution within the clusters that is normalized to reproduce the number of stars observed by Gaia in each cluster. To do this, it is necessary to define the mass range of the detected stars as integration limits of Eq.~\eqref{eq:IMF_kN}. We inferred these values from available observations as described in Sec.~\ref{sec:stars}, relying on the measured G-band stellar magnitudes, and converting them into bolometric magnitudes first, to then obtain stellar absolute luminosities after correction for the G-band extinction along the line of sight. We finally adopted the mass-luminosity relation in Eq.~\eqref{eqn:Lstar} relatively to the MS evolution to obtain stellar masses. The highest magnitude star per cluster hence determines the minimum stellar mass $M^*_{\rm min}$, while the lowest magnitude star determines the maximum stellar mass observed $M^*_{\rm max}$. In reality, the magnitude distribution of stars within clusters does not show a sharp drop at the highest magnitude, but it rather appears as a peaked distribution, whose mode is determined by the limiting magnitude of the observation in the given direction. Hence, we adopted the mode of the magnitude distribution rather that its highest value to infer the minimum stellar mass, shown in Fig.~\ref{fig:mmin}. The maximum stellar mass per cluster is in turn represented in Fig.~\ref{fig:mmax}, its highest value amounting to $\sim 18 M_\odot$. The scatter plot among the two is shown in Fig.~\ref{fig:smm}.

After fixing $M_{\rm min}^*$ and $M_{\rm max}^*$, the normalization constant $k$ of the cluster stellar mass function was obtained. We then determined the expected cluster masses and luminosities by extending the mass range of stars populating these systems, as suggested by theoretical arguments rather than being limited by instrument detection capabilities. To do this, we simply extended the integration mass range between $0.08 \, M_\odot$ and a maximum value, determined by either the cluster mass or its age, whichever yielded the lower value.

\begin{figure}
\centering
\includegraphics[width=0.45\textwidth]{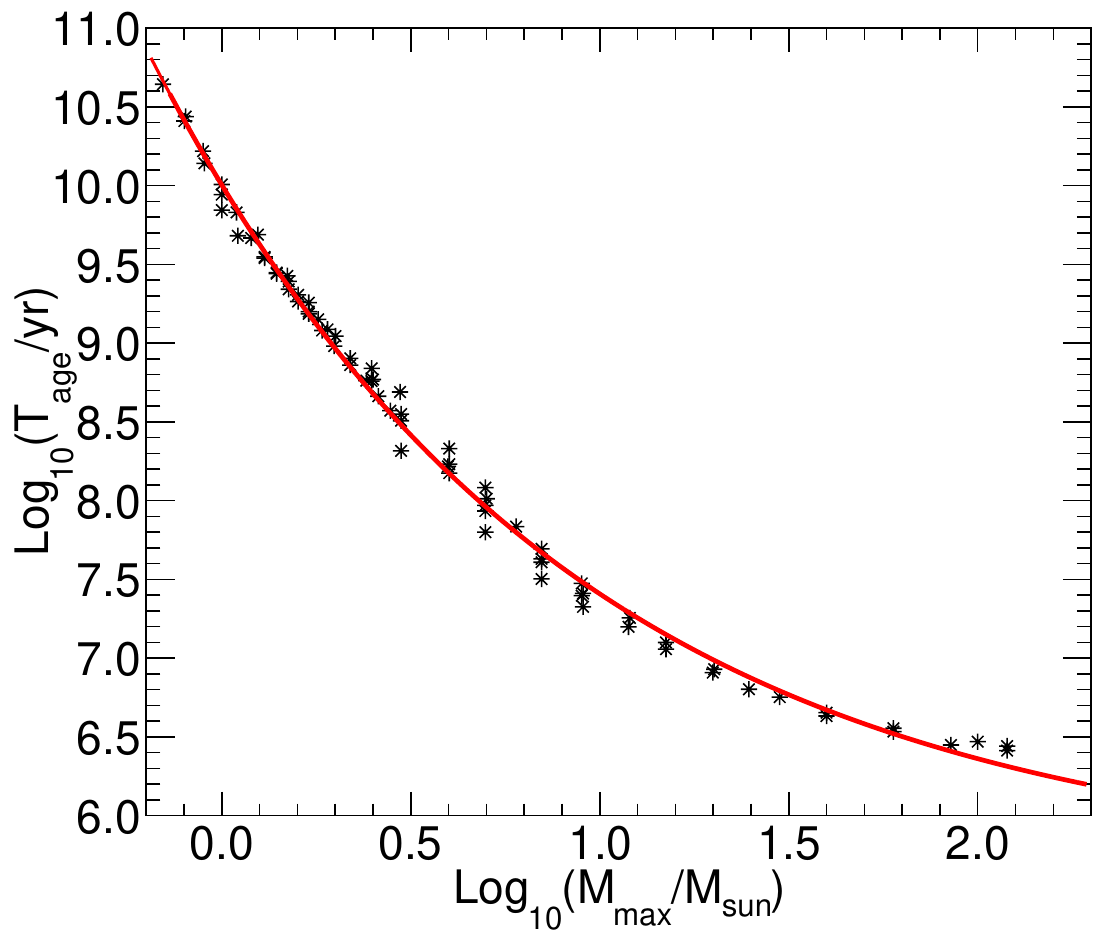}
\caption{Mass-age relation for main-sequence stars with data from \citet{buzzoni2002}. The functional form of the red line, obtained through a minimum residual technique, is provided in Eq.~\eqref{eq:param}.}
\label{fig:buzz}
\end{figure}

Older clusters are expected to host less massive stars than younger clusters because of the occurrence of SN explosions. The stellar lifetimes depend inversely on their mass. {\bf A}f $10 \, M_\odot$ star will explode in an SN after about 10~Myr, while a $100 \, M_\odot$ star will explode after only $\sim1$~Myr. We parameterized this relation through Eq.~\eqref{eq:param}, which we fit to stellar evolution data from \citet{buzzoni2002} in the mass range $-0.2 \leq \log_{10} [M/M_\odot] \leq 2.1$, as shown in Fig.~\ref{fig:buzz}. The result is shown in Fig.~\ref{fig:buzz}, where the red curve was obtained for the following set of values: $A=4.35 \pm 0.07$, $B=1.30 \pm 0.04$, and $C=1.101 \pm 0.035$, with a $\chi^2$ of 0.5. 

\begin{figure}
\centering
\includegraphics[width=0.45\textwidth]{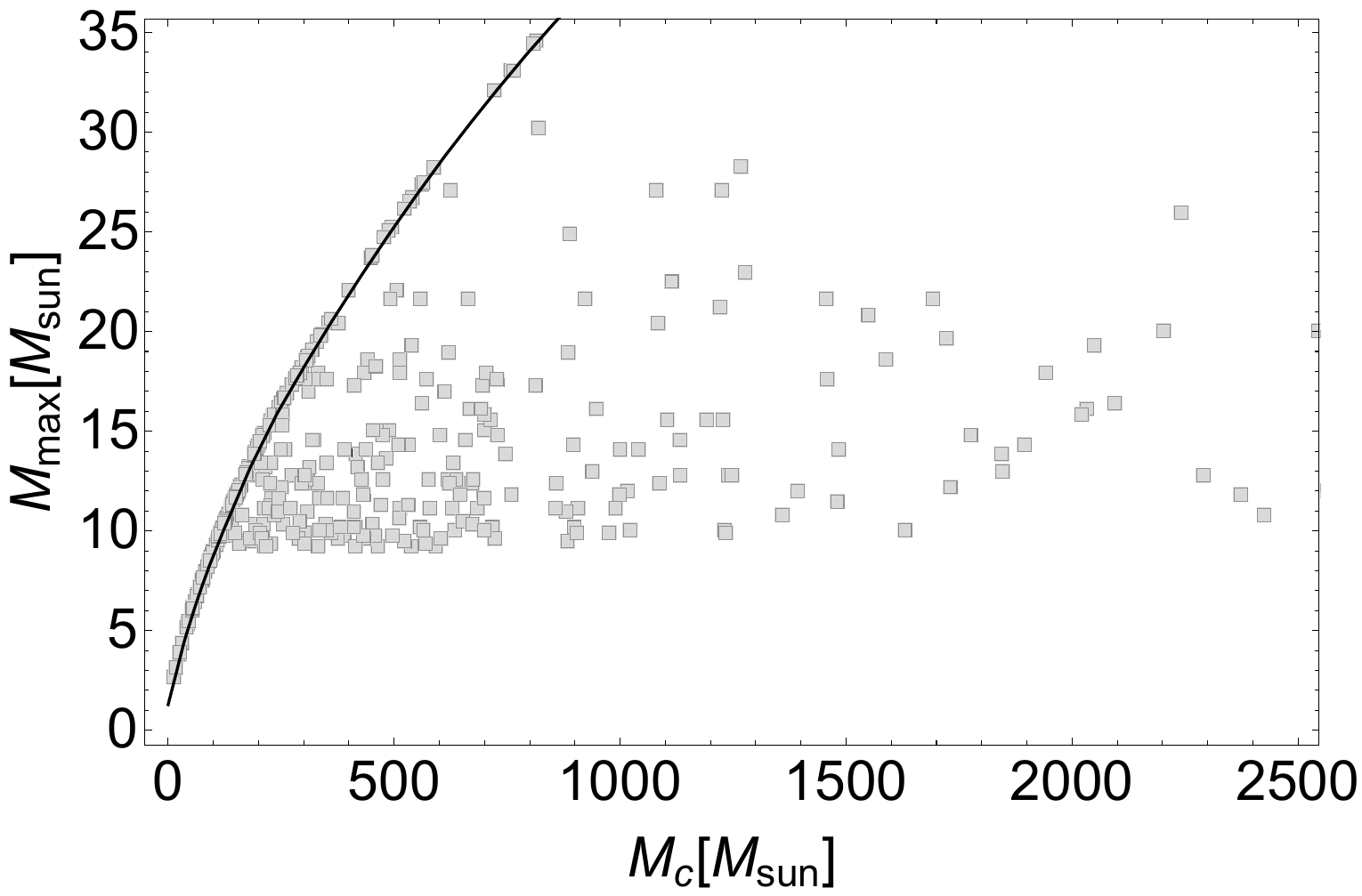}
\caption{Maximum stellar masses $M_{\rm max}$ as a function of the seed cluster masses, shown on the abscissa. The solid line represents the numerical solution we derived following \citet{Weidner-Kroupa:2006}. The gray markers refer to the selected young cluster sample.}
\label{fig:weid}
\end{figure}

For the effects of the parent cluster mass on the maximum stellar mass that might form within it, we followed the description from \citet{Weidner-Kroupa:2006}, represented in Fig.~\ref{fig:weid} as the solid black line. Because of the non{\bf -}linearity intrinsic to the application of this constraint, which would require knowing the cluster mass to determine the maximum stellar mass hosted, we implemented an iterative technique that took as input the seed cluster mass, evaluated with a maximum stellar mass limited by either the cluster age \citep{buzzoni2002} or by $150 \, M_\odot$. From this seed mass, we computed the expected $M_{\rm max}$ following \citet{Weidner-Kroupa:2006}. For all clusters of our sample for which the $M_{\rm max}$ previously evaluated through the age-limited maximum stellar mass was higher than the masses evaluated based on the seed cluster mass, we first corrected $M_{\rm max}$ with the latter value. As a consequence, we computed the updated cluster mass value $M_{\rm c}$ from Eq.~\eqref{eq:mc}, which would now correspond to a different maximum stellar mass according to \citet{Weidner-Kroupa:2006}. We proceeded with these iterations until convergence was achieved on the cluster mass values. The final $M_{\rm max}$ versus $M_{\rm c}$ values are shown as dots in Fig.~\ref{fig:weid} for the entire young cluster sample we analyzed. The markers closely following the theoretical line refer to the clusters for which the iterative procedure was effective. As visible, the parent cluster mass strongly constraints the maximum stellar mass that will be formed in low-mass clusters.

\section{Energetic considerations about the contributions of SNe and stellar winds}
\label{sec:appB}
\begin{figure}
\centering
\includegraphics[width=0.48\textwidth]{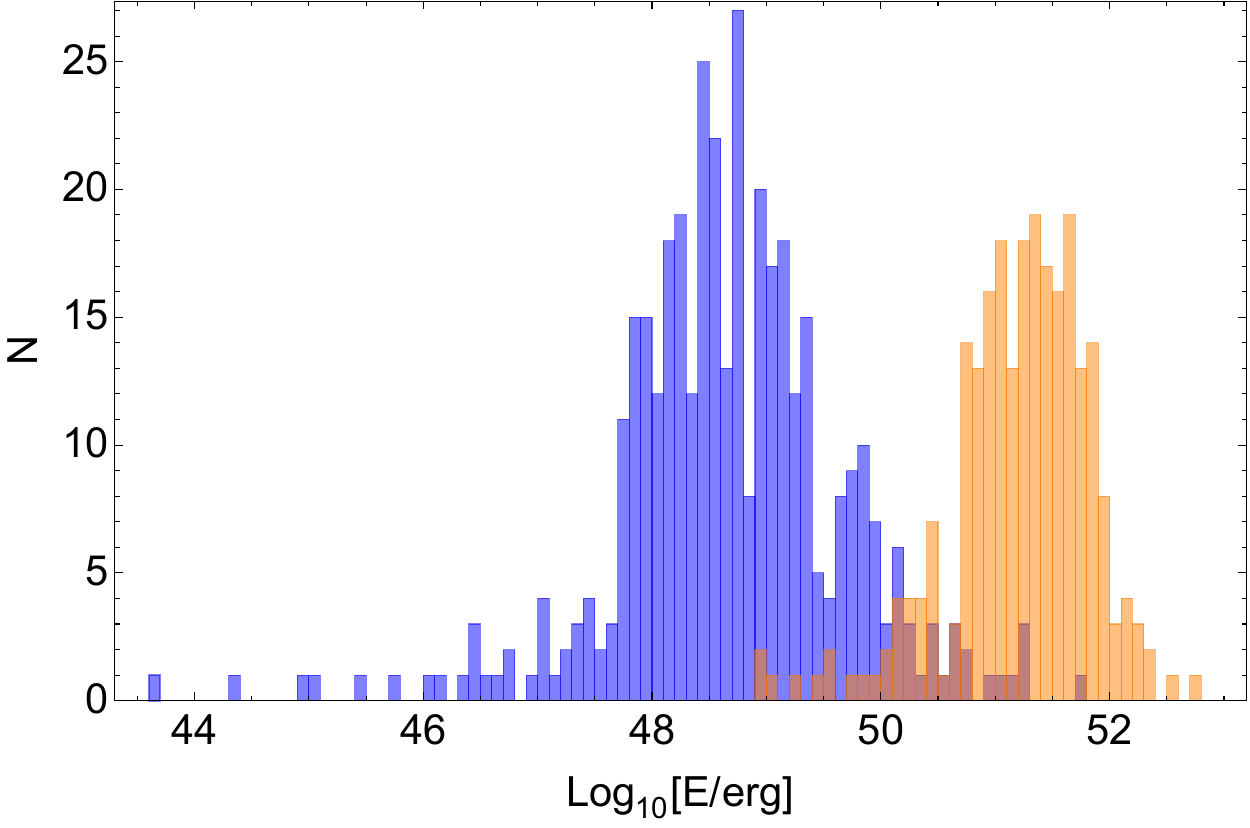}
\caption{Comparison of the expected energetics injected by cluster winds (blue histogram) and SN explosions (red histogram) in the selected sample of young stellar clusters.}
\label{fig:sne}
\end{figure}

The methods developed here for estimating the mass and wind luminosity of star clusters further allowed us to infer the temporal evolution of the stellar population of which individual stellar clusters are composed. Namely, we can trace the number of stars that have undergone SN explosions since cluster formation by computing the integral of the mass distribution function between the maximum stellar mass ever hosted by the cluster $M_{\rm max,0}$, as dictated by the relation to the parent cluster mass, and the maximum mass expected at the current age $M_{\rm max} (T_{\rm age})$ as
\begin{linenomath}
\begin{equation}
  N_{\rm SN} = \int_{M_{\rm max}(T_{\rm age})}^{M_{\rm max,0}} \xi(M) \, dM \, .
\end{equation}
\end{linenomath}
By assuming an energy release of $E_{\rm SN}=10^{51}$~erg per SN, we can compute the total energetics channeled into SNe explosions as $E^{\rm tot}_{\rm SN} = N_{\rm SN} E_{\rm SN}$. For the total energetics transferred to stellar winds, we estimate
\begin{linenomath}
\begin{equation}
\label{eq:Ewind}
 E_{\rm w,c} = \frac{1}{2} \int_{M_{\rm min}}^{M_{\rm max}} \dot{M}_{\rm s}(M) \, v^2_{\rm w,c}(M) \, \tau_{\rm s}(M) \, \xi(M) \, dM \, ,
\end{equation}
\end{linenomath}
where $\tau_{\rm s}(M)$ represents the MS duration for a star of mass $M$, as shown in Fig.~\ref{fig:buzz}. The total energy transferred into SNe and winds resulting from this computation is shown in Fig.~\ref{fig:sne}. The obtained median values amount to $7 \times 10^{50}$~erg and $5 \times 10^{48}$~erg for SNe and winds, respectively. This illustrates the relevance of SNe in the energetics considerations of stellar clusters, and therefore also the relevance for particle acceleration and related gamma-ray emission. We note, however, that Eq.~\eqref{eq:Ewind} is expected to be higher when the contribution of WRs is included. Nonetheless, this simplified estimate already suggests that the energy injected into winds throughout the cluster history is higher than $10^{51}$~erg for some systems, namely Wd1, Danks1, Danks2, Patchick94, UBC344, and NGC3603.

\section{Content of accompanying data}
\label{sec:appC}
Online material supplements the content of this manuscript: a text file is provided in a machine-readable format, whose content corresponds to Tab.~\ref{tab:cluster_parameters}.

\clearpage
\onecolumn

\newpage
\begin{table}
\caption{Cluster name (Gaia DR2 label), age in Myr, estimated mass in $M_\odot$, mass-loss rate in $M_\odot$/yr, wind luminosity in erg/s, normalization constant $k$ of its stellar mass function according to Eq.~\eqref{eq:imf}, and the inferred minimum and maximum stellar mass observed by Gaia as described in Sec.~\ref{sec:stars}, $M^*_{\rm min}$ and $M^*_{\rm max}$ respectively.}
\label{tab:cluster_parameters}
\end{table}
\vspace{-5cm}
\begin{longtable}{lccccccc}
\centering
Name & $T_{\rm age}$ [Myr] & $M_{\rm c}$ [$M_\odot$] & $\dot{M}_{\rm c}$ [$M_\odot$/yr] & $L_{\rm w,c}$ [\rm erg/s] & k & $M^*_{\rm min}$ [$M_\odot$] & $M^*_{\rm max}$ [$M_\odot$] \\
\hline\hline
\endfirsthead
\multicolumn{8}{l}{\footnotesize\itshape Table continues from the previous page} \\
Name & Age [Myr] & $M_{\rm c}$ [$M_\odot$] & $\dot{M}_{\rm c}$ [$M_\odot$/yr] & $L_{\rm w,c}$ [\rm erg/s] & k & $M^*_{\rm min}$ [$M_\odot$] & $M^*_{\rm max}$ [$M_\odot$] \\
\hline\hline
\endhead
\multicolumn{8}{r}{\footnotesize\itshape Table continues in the next page} \\
\endfoot
\multicolumn{8}{r}{\footnotesize\itshape Table ends from the previous page} \\
\endlastfoot
ASCC\char`_107 & 17.0 & 216.9 & $6.67\times10^{-9}$ & $1.39\times10^{34}$ & 3093.9 & 1.25 & 6.07 \\
ASCC\char`_127 & 18.2 & 108.7 & $1.14\times10^{-9}$ & $2.09\times10^{33}$ & 1616.2 & 0.43 & 2.78 \\
ASCC\char`_16 & 13.5 & 161.4 & $3.91\times10^{-9}$ & $7.94\times10^{33}$ & 2323.5 & 0.42 & 3.10 \\
ASCC\char`_19 & 10.5 & 181.2 & $5.63\times10^{-9}$ & $1.18\times10^{34}$ & 2584.6 & 0.96 & 4.28 \\
ASCC\char`_21 & 8.9 & 45.6 & $5.39\times10^{-11}$ & $5.90\times10^{31}$ & 736.6 & 0.36 & 3.28 \\
ASCC\char`_32 & 25.1 & 559.6 & $7.83\times10^{-9}$ & $1.49\times10^{34}$ & 8226.4 & 1.04 & 3.25 \\
ASCC\char`_79 & 18.6 & 670.9 & $1.71\times10^{-8}$ & $3.49\times10^{34}$ & 9637.9 & 1.53 & 4.54 \\
ASCC\char`_9 & 29.5 & 594.8 & $6.04\times10^{-9}$ & $1.11\times10^{34}$ & 8852.6 & 2.71 & 6.90 \\
Alessi\char`_18 & 5.1 & 310.7 & $2.63\times10^{-8}$ & $6.17\times10^{34}$ & 4278.7 & 2.25 & 4.07 \\
Alessi\char`_19 & 25.7 & 125.9 & $1.68\times10^{-9}$ & $3.19\times10^{33}$ & 1853.3 & 0.88 & 5.26 \\
Alessi\char`_20 & 9.3 & 130.5 & $2.04\times10^{-9}$ & $3.95\times10^{33}$ & 1909.2 & 1.07 & 3.36 \\
Alessi\char`_43 & 11.5 & 574.3 & $4.10\times10^{-8}$ & $9.41\times10^{34}$ & 7955.4 & 1.12 & 6.58 \\
Alessi\char`_59 & 10.0 & 201.1 & $7.81\times10^{-9}$ & $1.67\times10^{34}$ & 2844.9 & 2.69 & 3.47 \\
Alessi\char`_Teutsch\char`_5 & 18.6 & 316.0 & $8.05\times10^{-9}$ & $1.64\times10^{34}$ & 4539.8 & 1.08 & 3.70 \\
Andrews\char`_Lindsay\char`_5 & 1.0 & 1029.7 & $6.79\times10^{-7}$ & $2.02\times10^{36}$ & 13356.9 & 3.06 & 5.62 \\
Antalova\char`_2 & 12.0 & 143.4 & $2.73\times10^{-9}$ & $5.39\times10^{33}$ & 2083.3 & 1.54 & 2.17 \\
Aveni\char`_Hunter\char`_1 & 10.5 & 93.3 & $7.08\times10^{-10}$ & $1.25\times10^{33}$ & 1404.4 & 1.22 & 2.80 \\
BDSB30 & 3.5 & 43.2 & $4.53\times10^{-11}$ & $4.68\times10^{31}$ & 702.1 & 1.63 & 3.18 \\
BDSB91 & 14.1 & 136.1 & $2.32\times10^{-9}$ & $4.52\times10^{33}$ & 1985.4 & 1.51 & 2.54 \\
BDSB93 & 15.8 & 61.9 & $1.64\times10^{-10}$ & $2.39\times10^{32}$ & 969.6 & 1.41 & 5.23 \\
BDSB96 & 7.9 & 99.3 & $8.67\times10^{-10}$ & $1.56\times10^{33}$ & 1487.4 & 1.33 & 5.50 \\
BH\char`_121 & 2.6 & 1251.8 & $1.07\times10^{-6}$ & $3.27\times10^{36}$ & 16117.6 & 1.80 & 12.30 \\
BH\char`_150 & 21.9 & 581.7 & $1.07\times10^{-8}$ & $2.11\times10^{34}$ & 8459.6 & 2.82 & 7.29 \\
BH\char`_151 & 15.8 & 417.5 & $1.48\times10^{-8}$ & $3.15\times10^{34}$ & 5925.2 & 3.43 & 10.21 \\
BH\char`_205 & 6.2 & 452.3 & $7.56\times10^{-8}$ & $1.91\times10^{35}$ & 6099.2 & 1.84 & 4.76 \\
BH\char`_221 & 10.7 & 515.5 & $4.30\times10^{-8}$ & $1.01\times10^{35}$ & 7103.6 & 1.29 & 6.64 \\
BH\char`_23 & 28.8 & 79.3 & $4.03\times10^{-10}$ & $6.71\times10^{32}$ & 1211.8 & 0.43 & 5.71 \\
BH\char`_245 & 4.7 & 543.2 & $1.27\times10^{-7}$ & $3.33\times10^{35}$ & 7254.8 & 4.07 & 9.08 \\
BH\char`_54 & 8.7 & 508.3 & $6.92\times10^{-8}$ & $1.71\times10^{35}$ & 6897.3 & 2.84 & 7.42 \\
BH\char`_56 & 17.8 & 311.4 & $8.71\times10^{-9}$ & $1.80\times10^{34}$ & 4457.7 & 1.05 & 4.47 \\
Basel\char`_18 & 27.5 & 211.8 & $2.47\times10^{-9}$ & $4.60\times10^{33}$ & 3136.0 & 1.26 & 5.09 \\
Berkeley\char`_14A & 9.3 & 97.0 & $8.04\times10^{-10}$ & $1.43\times10^{33}$ & 1456.0 & 1.41 & 1.60 \\
Berkeley\char`_15 & 18.6 & 1089.5 & $2.77\times10^{-8}$ & $5.67\times10^{34}$ & 15651.0 & 1.71 & 4.62 \\
Berkeley\char`_4 & 21.4 & 473.4 & $9.12\times10^{-9}$ & $1.81\times10^{34}$ & 6872.9 & 2.55 & 8.95 \\
Berkeley\char`_47 & 21.9 & 631.2 & $1.16\times10^{-8}$ & $2.29\times10^{34}$ & 9180.3 & 2.69 & 6.86 \\
Berkeley\char`_59 & 1.3 & 1323.8 & $1.21\times10^{-6}$ & $3.75\times10^{36}$ & 17007.6 & 2.51 & 9.58 \\
Berkeley\char`_62 & 14.8 & 660.5 & $2.71\times10^{-8}$ & $5.85\times10^{34}$ & 9324.2 & 1.63 & 5.78 \\
Berkeley\char`_63 & 18.2 & 621.8 & $1.66\times10^{-8}$ & $3.41\times10^{34}$ & 8917.3 & 2.53 & 6.77 \\
Berkeley\char`_86 & 11.0 & 180.1 & $1.20\times10^{-5}$ & $9.25\times10^{36}$ & 2569.0 & 1.74 & 5.16 \\
Berkeley\char`_87 & 8.3 & 1279.1 & $1.36\times10^{-5}$ & $2.28\times10^{37}$ & 17298.2 & 2.62 & 10.86 \\
Berkeley\char`_90 & 5.1 & 817.2 & $3.82\times10^{-7}$ & $1.09\times10^{36}$ & 10704.6 & 2.59 & 13.93 \\
Berkeley\char`_94 & 15.5 & 440.0 & $1.64\times10^{-8}$ & $3.50\times10^{34}$ & 6233.8 & 2.31 & 8.35 \\
Berkeley\char`_95 & 12.0 & 313.0 & $2.01\times10^{-8}$ & $4.57\times10^{34}$ & 4349.8 & 2.28 & 7.63 \\
Berkeley\char`_97 & 15.1 & 534.0 & $2.09\times10^{-8}$ & $4.48\times10^{34}$ & 7552.1 & 1.70 & 7.22 \\
Bica\char`_3 & 15.8 & 749.0 & $2.66\times10^{-8}$ & $5.64\times10^{34}$ & 10629.5 & 3.75 & 7.47 \\
Biurakan\char`_2 & 9.3 & 297.5 & $2.32\times10^{-8}$ & $5.39\times10^{34}$ & 4108.6 & 1.48 & 6.42 \\
Bochum\char`_11 & 6.3 & 357.9 & $3.92\times10^{-8}$ & $9.44\times10^{34}$ & 4890.0 & 2.12 & 10.95 \\
Bochum\char`_13 & 8.9 & 494.2 & $6.36\times10^{-8}$ & $1.56\times10^{35}$ & 6718.0 & 2.36 & 9.91 \\
Bochum\char`_6 & 6.5 & 497.9 & $9.91\times10^{-8}$ & $2.56\times10^{35}$ & 6679.9 & 1.68 & 7.57 \\
COIN-Gaia\char`_16 & 25.7 & 135.3 & $1.81\times10^{-9}$ & $3.43\times10^{33}$ & 1992.5 & 1.49 & 3.49 \\
COIN-Gaia\char`_21 & 28.2 & 169.3 & $1.89\times10^{-9}$ & $3.50\times10^{33}$ & 2511.4 & 1.40 & 4.08 \\
Collinder\char`_106 & 24.5 & 454.7 & $6.66\times10^{-9}$ & $1.28\times10^{34}$ & 6671.6 & 1.42 & 3.87 \\
Collinder\char`_107 & 18.2 & 579.0 & $1.55\times10^{-8}$ & $3.17\times10^{34}$ & 8303.4 & 1.43 & 6.01 \\
Collinder\char`_132 & 24.5 & 196.2 & $2.87\times10^{-9}$ & $5.51\times10^{33}$ & 2879.6 & 0.83 & 3.37 \\
Collinder\char`_135 & 26.3 & 305.4 & $3.90\times10^{-9}$ & $7.35\times10^{33}$ & 4505.3 & 0.38 & 4.40 \\
Collinder\char`_140 & 26.9 & 134.8 & $1.64\times10^{-9}$ & $3.08\times10^{33}$ & 1991.7 & 0.40 & 5.12 \\
Collinder\char`_197 & 14.1 & 491.2 & $2.22\times10^{-8}$ & $4.85\times10^{34}$ & 6910.5 & 1.20 & 7.21 \\
Collinder\char`_205 & 4.6 & 563.6 & $1.41\times10^{-7}$ & $3.72\times10^{35}$ & 7513.7 & 1.58 & 5.89 \\
Collinder\char`_419 & 19.5 & 143.4 & $2.71\times10^{-9}$ & $5.36\times10^{33}$ & 2083.3 & 1.40 & 9.94 \\
Collinder\char`_469 & 9.5 & 379.6 & $4.14\times10^{-8}$ & $9.98\times10^{34}$ & 5186.6 & 2.56 & 5.00 \\
Collinder\char`_69 & 12.6 & 564.1 & $3.28\times10^{-8}$ & $7.36\times10^{34}$ & 7868.1 & 1.00 & 3.75 \\
Collinder\char`_95 & 19.1 & 156.8 & $3.56\times10^{-9}$ & $7.19\times10^{33}$ & 2262.1 & 1.26 & 5.72 \\
Czernik\char`_31 & 29.5 & 466.9 & $4.74\times10^{-9}$ & $8.70\times10^{33}$ & 6949.1 & 1.43 & 4.86 \\
Czernik\char`_40 & 25.1 & 901.0 & $1.26\times10^{-8}$ & $2.40\times10^{34}$ & 13244.7 & 2.68 & 5.07 \\
Czernik\char`_41 & 12.9 & 2035.1 & $1.12\times10^{-7}$ & $2.51\times10^{35}$ & 28432.7 & 2.92 & 7.93 \\
Czernik\char`_6 & 28.2 & 193.1 & $2.15\times10^{-9}$ & $3.99\times10^{33}$ & 2863.7 & 2.40 & 4.51 \\
DB2001\char`_22 & 17.2 & 164.4 & $4.12\times10^{-9}$ & $8.40\times10^{33}$ & 2362.8 & 2.87 & 8.26 \\
DBSB\char`_101 & 11.2 & 334.5 & $2.51\times10^{-8}$ & $5.81\times10^{34}$ & 4624.7 & 1.77 & 4.06 \\
DBSB\char`_104 & 26.9 & 67.4 & $2.22\times10^{-10}$ & $3.41\times10^{32}$ & 1047.6 & 1.17 & 2.73 \\
DBSB\char`_21 & 9.5 & 119.9 & $1.57\times10^{-9}$ & $2.96\times10^{33}$ & 1766.8 & 2.39 & 4.18 \\
DBSB\char`_3 & 3.8 & 331.7 & $3.17\times10^{-8}$ & $7.53\times10^{34}$ & 4550.7 & 1.90 & 5.35 \\
DBSB\char`_43 & 24.5 & 256.9 & $3.76\times10^{-9}$ & $7.21\times10^{33}$ & 3769.1 & 1.88 & 4.82 \\
DBSB\char`_60 & 5.6 & 252.5 & $1.48\times10^{-8}$ & $3.31\times10^{34}$ & 3521.3 & 2.03 & 5.58 \\
DC\char`_5 & 1.1 & 1885.2 & $2.42\times10^{-6}$ & $7.87\times10^{36}$ & 23961.6 & 2.74 & 5.35 \\
Danks\char`_1 & 1.0 & 4358.2 & $8.53\times10^{-5}$ & $7.76\times10^{37}$ & 54214.2 & 5.57 & 7.13 \\
Danks\char`_2 & 2.0 & 7811.8 & $3.85\times10^{-5}$ & $7.01\times10^{37}$ & 96204.2 & 5.37 & 6.03 \\
Dias\char`_5 & 9.5 & 297.9 & $2.33\times10^{-8}$ & $5.41\times10^{34}$ & 4113.7 & 1.38 & 3.38 \\
Dolidze\char`_11 & 22.4 & 310.7 & $5.46\times10^{-9}$ & $1.07\times10^{34}$ & 4526.3 & 2.34 & 3.93 \\
Dolidze\char`_16 & 16.6 & 215.8 & $6.96\times10^{-9}$ & $1.46\times10^{34}$ & 3072.6 & 1.39 & 4.16 \\
Dolidze\char`_3 & 8.9 & 131.2 & $1.34\times10^{-5}$ & $2.31\times10^{37}$ & 1918.9 & 1.68 & 3.94 \\
Dolidze\char`_53 & 6.9 & 194.5 & $7.05\times10^{-9}$ & $1.50\times10^{34}$ & 2757.3 & 2.31 & 4.29 \\
Dolidze\char`_8 & 15.5 & 77.9 & $3.82\times10^{-10}$ & $6.33\times10^{32}$ & 1192.1 & 1.43 & 4.47 \\
FSR\char`_0158 & 10.2 & 150.2 & $3.15\times10^{-9}$ & $6.30\times10^{33}$ & 2174.1 & 2.59 & 3.60 \\
FSR\char`_0165 & 15.5 & 1043.0 & $3.89\times10^{-8}$ & $8.29\times10^{34}$ & 14775.7 & 3.27 & 5.53 \\
FSR\char`_0198 & 4.7 & 451.0 & $7.51\times10^{-8}$ & $1.90\times10^{35}$ & 6082.4 & 1.48 & 5.83 \\
FSR\char`_0398 & 18.2 & 211.6 & $5.65\times10^{-9}$ & $1.16\times10^{34}$ & 3034.3 & 1.32 & 4.29 \\
FSR\char`_0451 & 13.5 & 1230.5 & $6.15\times10^{-8}$ & $1.36\times10^{35}$ & 17251.4 & 1.63 & 9.87 \\
FSR\char`_0534 & 25.7 & 567.3 & $7.58\times10^{-9}$ & $1.44\times10^{34}$ & 8354.1 & 2.32 & 5.67 \\
FSR\char`_0551 & 13.5 & 203.7 & $8.10\times10^{-9}$ & $1.74\times10^{34}$ & 2878.7 & 1.13 & 3.37 \\
FSR\char`_0826 & 4.6 & 229.3 & $1.13\times10^{-8}$ & $2.50\times10^{34}$ & 3215.7 & 1.98 & 6.26 \\
FSR\char`_0833 & 16.6 & 230.3 & $7.43\times10^{-9}$ & $1.56\times10^{34}$ & 3279.8 & 2.31 & 6.42 \\
FSR\char`_0904 & 21.9 & 515.7 & $9.49\times10^{-9}$ & $1.87\times10^{34}$ & 7500.7 & 1.58 & 11.17 \\
FSR\char`_1117 & 15.5 & 47.7 & $6.30\times10^{-11}$ & $7.26\times10^{31}$ & 767.4 & 1.21 & 3.14 \\
FSR\char`_1297 & 5.0 & 147.5 & $3.00\times10^{-9}$ & $5.97\times10^{33}$ & 2136.4 & 1.72 & 2.84 \\
FSR\char`_1335 & 11.2 & 151.0 & $3.20\times10^{-9}$ & $6.40\times10^{33}$ & 2185.0 & 2.20 & 4.99 \\
FSR\char`_1342 & 25.7 & 636.8 & $8.51\times10^{-9}$ & $1.61\times10^{34}$ & 9377.4 & 2.03 & 4.51 \\
FSR\char`_1352 & 3.5 & 145.9 & $2.92\times10^{-9}$ & $5.80\times10^{33}$ & 2115.6 & 1.89 & 4.85 \\
FSR\char`_1435 & 1.0 & 401.7 & $5.46\times10^{-8}$ & $1.35\times10^{35}$ & 5451.6 & 2.38 & 5.56 \\
Graham\char`_1 & 15.5 & 261.8 & $9.75\times10^{-9}$ & $2.08\times10^{34}$ & 3708.6 & 3.55 & 7.65 \\
Gulliver\char`_10 & 9.3 & 79.3 & $4.12\times10^{-10}$ & $6.87\times10^{32}$ & 1211.8 & 1.03 & 2.76 \\
Gulliver\char`_2 & 6.6 & 229.8 & $1.14\times10^{-8}$ & $2.51\times10^{34}$ & 3222.2 & 1.15 & 7.25 \\
Gulliver\char`_29 & 10.7 & 1590.8 & $1.33\times10^{-7}$ & $3.10\times10^{35}$ & 21921.7 & 1.35 & 5.18 \\
Gulliver\char`_46 & 11.7 & 698.8 & $4.74\times10^{-8}$ & $1.08\times10^{35}$ & 9695.9 & 2.66 & 13.71 \\
Gulliver\char`_48 & 18.2 & 640.0 & $1.71\times10^{-8}$ & $3.51\times10^{34}$ & 9178.5 & 1.96 & 3.22 \\
Gulliver\char`_6 & 16.6 & 353.0 & $1.14\times10^{-8}$ & $2.39\times10^{34}$ & 5027.1 & 0.99 & 4.65 \\
Gulliver\char`_8 & 21.4 & 102.7 & $9.51\times10^{-10}$ & $1.72\times10^{33}$ & 1534.7 & 1.24 & 5.89 \\
Gulliver\char`_9 & 17.8 & 276.3 & $7.73\times10^{-9}$ & $1.60\times10^{34}$ & 3956.0 & 0.45 & 5.05 \\
Haffner\char`_13 & 20.4 & 231.6 & $4.90\times10^{-9}$ & $9.79\times10^{33}$ & 3350.8 & 0.42 & 5.23 \\
Haffner\char`_15 & 25.7 & 1233.1 & $1.65\times10^{-8}$ & $3.12\times10^{34}$ & 18158.1 & 2.01 & 9.57 \\
Haffner\char`_18 & 14.1 & 702.3 & $3.18\times10^{-8}$ & $6.94\times10^{34}$ & 9880.4 & 2.83 & 6.19 \\
Havlen\char`_Moffat\char`_1 & 8.9 & 666.1 & $8.57\times10^{-8}$ & $2.10\times10^{35}$ & 9054.4 & 4.24 & 11.11 \\
Hogg\char`_10 & 6.1 & 120.1 & $1.58\times10^{-9}$ & $2.99\times10^{33}$ & 1768.8 & 1.97 & 3.35 \\
Hogg\char`_15 & 2.2 & 1064.0 & $1.27\times10^{-5}$ & $8.84\times10^{36}$ & 13784.9 & 2.84 & 10.66 \\
Hogg\char`_19 & 18.6 & 333.2 & $8.48\times10^{-9}$ & $1.73\times10^{34}$ & 4786.4 & 1.95 & 5.28 \\
IC\char`_1396 & 12.0 & 614.2 & $3.95\times10^{-8}$ & $8.97\times10^{34}$ & 8537.3 & 1.10 & 7.10 \\
IC\char`_1442 & 19.5 & 1018.6 & $2.36\times10^{-8}$ & $4.78\times10^{34}$ & 14684.6 & 2.41 & 4.99 \\
IC\char`_1590 & 6.2 & 567.2 & $1.43\times10^{-7}$ & $3.79\times10^{35}$ & 7558.5 & 1.97 & 9.38 \\
IC\char`_1805 & 7.6 & 891.2 & $1.71\times10^{-7}$ & $4.39\times10^{35}$ & 11971.2 & 2.34 & 13.82 \\
IC\char`_1848 & 15.8 & 1846.8 & $6.56\times10^{-8}$ & $1.39\times10^{35}$ & 26207.5 & 1.58 & 8.28 \\
IC\char`_2391 & 28.8 & 159.7 & $1.70\times10^{-9}$ & $3.13\times10^{33}$ & 2372.1 & 0.28 & 3.43 \\
IC\char`_2395 & 20.4 & 388.9 & $8.22\times10^{-9}$ & $1.64\times10^{34}$ & 5626.5 & 0.51 & 5.85 \\
IC\char`_2581 & 10.2 & 541.4 & $5.03\times10^{-8}$ & $1.19\times10^{35}$ & 7435.0 & 1.71 & 9.19 \\
IC\char`_348 & 11.7 & 101.3 & $9.17\times10^{-10}$ & $1.66\times10^{33}$ & 1514.8 & 1.52 & 2.53 \\
IC\char`_4996 & 11.2 & 436.6 & $3.28\times10^{-8}$ & $7.58\times10^{34}$ & 6037.4 & 1.55 & 7.56 \\
IC\char`_5146 & 1.0 & 77.0 & $3.79\times10^{-10}$ & $6.27\times10^{32}$ & 1178.2 & 1.29 & 5.57 \\
Juchert\char`_18 & 24.5 & 213.3 & $3.12\times10^{-9}$ & $5.98\times10^{33}$ & 3129.2 & 1.72 & 4.42 \\
Juchert\char`_19 & 8.1 & 215.0 & $9.47\times10^{-9}$ & $2.06\times10^{34}$ & 3027.7 & 2.52 & 9.33 \\
Juchert\char`_3 & 1.0 & 2157.7 & $3.16\times10^{-6}$ & $1.05\times10^{37}$ & 27306.8 & 4.95 & 5.42 \\
Juchert\char`_9 & 11.5 & 334.6 & $2.39\times10^{-8}$ & $5.48\times10^{34}$ & 4635.2 & 1.97 & 6.03 \\
King\char`_10 & 17.4 & 1848.8 & $5.42\times10^{-8}$ & $1.13\times10^{35}$ & 26419.9 & 2.33 & 9.21 \\
King\char`_12 & 18.6 & 675.3 & $1.72\times10^{-8}$ & $3.51\times10^{34}$ & 9700.9 & 2.65 & 7.35 \\
Koposov\char`_53 & 16.2 & 485.1 & $1.64\times10^{-8}$ & $3.46\times10^{34}$ & 6895.8 & 2.25 & 4.96 \\
Kronberger\char`_1 & 6.0 & 321.2 & $2.89\times10^{-8}$ & $6.81\times10^{34}$ & 4415.6 & 2.11 & 6.07 \\
LDN\char`_988e & 4.7 & 14.8 & $1.02\times10^{-12}$ & $2.79\times10^{29}$ & 276.7 & 1.70 & 2.97 \\
LP\char`_403 & 26.3 & 978.4 & $1.25\times10^{-8}$ & $2.36\times10^{34}$ & 14433.7 & 1.64 & 4.73 \\
L\char`_1641S & 4.2 & 118.4 & $1.52\times10^{-9}$ & $2.87\times10^{33}$ & 1746.1 & 1.93 & 3.24 \\
Lynga\char`_6 & 3.1 & 454.1 & $7.68\times10^{-8}$ & $1.95\times10^{35}$ & 6122.4 & 2.62 & 5.65 \\
Mamajek\char`_1 & 9.8 & 26.9 & $7.96\times10^{-12}$ & $4.47\times10^{30}$ & 462.4 & 1.09 & 2.68 \\
Markarian\char`_38 & 15.8 & 66.3 & $2.10\times10^{-10}$ & $3.21\times10^{32}$ & 1031.2 & 1.42 & 3.22 \\
Markarian\char`_50 & 11.5 & 307.3 & $1.20\times10^{-5}$ & $9.24\times10^{36}$ & 4256.0 & 1.66 & 5.59 \\
Mayer\char`_1 & 28.8 & 230.3 & $2.45\times10^{-9}$ & $4.52\times10^{33}$ & 3422.2 & 1.63 & 5.62 \\
Muzzio\char`_1 & 7.8 & 265.2 & $1.68\times10^{-8}$ & $3.82\times10^{34}$ & 3688.3 & 2.26 & 6.55 \\
NGC\char`_1444 & 9.8 & 132.3 & $2.13\times10^{-9}$ & $4.13\times10^{33}$ & 1934.0 & 1.86 & 5.55 \\
NGC\char`_1502 & 10.5 & 887.9 & $7.81\times10^{-8}$ & $1.84\times10^{35}$ & 12214.5 & 1.70 & 6.12 \\
NGC\char`_1579 & 3.8 & 234.9 & $1.21\times10^{-8}$ & $2.69\times10^{34}$ & 3289.7 & 2.24 & 2.79 \\
NGC\char`_1624 & 4.5 & 312.4 & $2.68\times10^{-8}$ & $6.28\times10^{34}$ & 4300.8 & 3.05 & 3.98 \\
NGC\char`_1893 & 4.4 & 1205.1 & $9.57\times10^{-7}$ & $2.91\times10^{36}$ & 15548.1 & 2.14 & 8.23 \\
NGC\char`_1960 & 28.2 & 886.1 & $9.87\times10^{-9}$ & $1.83\times10^{34}$ & 13142.7 & 1.18 & 5.90 \\
NGC\char`_1980 & 13.2 & 130.4 & $2.03\times10^{-9}$ & $3.92\times10^{33}$ & 1908.9 & 0.95 & 2.85 \\
NGC\char`_2129 & 16.6 & 466.7 & $1.51\times10^{-8}$ & $3.16\times10^{34}$ & 6646.3 & 1.63 & 12.97 \\
NGC\char`_2169 & 12.3 & 198.9 & $7.52\times10^{-9}$ & $1.61\times10^{34}$ & 2815.6 & 1.20 & 4.77 \\
NGC\char`_2183 & 17.4 & 72.9 & $3.00\times10^{-10}$ & $4.81\times10^{32}$ & 1123.2 & 1.00 & 2.91 \\
NGC\char`_2232 & 17.8 & 179.0 & $5.01\times10^{-9}$ & $1.03\times10^{34}$ & 2561.8 & 0.40 & 4.59 \\
NGC\char`_2244 & 12.6 & 2096.9 & $1.22\times10^{-7}$ & $2.73\times10^{35}$ & 29244.9 & 1.45 & 8.62 \\
NGC\char`_2264 & 27.5 & 291.5 & $3.40\times10^{-9}$ & $6.34\times10^{33}$ & 4316.0 & 1.03 & 4.22 \\
NGC\char`_2362 & 5.8 & 591.7 & $1.61\times10^{-7}$ & $4.30\times10^{35}$ & 7868.7 & 1.14 & 5.29 \\
NGC\char`_2367 & 14.1 & 194.0 & $6.94\times10^{-9}$ & $1.47\times10^{34}$ & 2751.6 & 1.53 & 7.33 \\
NGC\char`_2414 & 17.8 & 1242.5 & $3.48\times10^{-8}$ & $7.18\times10^{34}$ & 17786.3 & 2.20 & 6.35 \\
NGC\char`_2439 & 11.7 & 2711.8 & $1.84\times10^{-7}$ & $4.19\times10^{35}$ & 37625.5 & 1.42 & 17.13 \\
NGC\char`_2453 & 25.1 & 720.1 & $1.01\times10^{-8}$ & $1.92\times10^{34}$ & 10585.0 & 1.80 & 13.74 \\
NGC\char`_2571 & 26.9 & 460.0 & $5.61\times10^{-9}$ & $1.05\times10^{34}$ & 6798.4 & 0.99 & 5.82 \\
NGC\char`_2645 & 25.7 & 196.1 & $2.62\times10^{-9}$ & $4.97\times10^{33}$ & 2887.3 & 1.49 & 6.19 \\
NGC\char`_3105 & 21.9 & 909.0 & $1.67\times10^{-8}$ & $3.29\times10^{34}$ & 13220.9 & 2.94 & 11.39 \\
NGC\char`_3293 & 10.2 & 2051.5 & $1.90\times10^{-7}$ & $4.50\times10^{35}$ & 28173.4 & 1.52 & 14.35 \\
NGC\char`_3324 & 10.7 & 443.5 & $3.70\times10^{-8}$ & $8.65\times10^{34}$ & 6112.1 & 2.76 & 10.42 \\
NGC\char`_3572 & 4.8 & 542.8 & $1.26\times10^{-7}$ & $3.32\times10^{35}$ & 7249.3 & 1.51 & 7.38 \\
NGC\char`_3590 & 26.3 & 906.5 & $1.16\times10^{-8}$ & $2.18\times10^{34}$ & 13372.3 & 1.48 & 10.70 \\
NGC\char`_3603 & 1.0 & 4808.4 & $7.21\times10^{-5}$ & $4.60\times10^{37}$ & 59700.4 & 3.09 & 13.31 \\
NGC\char`_366 & 25.7 & 1024.4 & $1.37\times10^{-8}$ & $2.60\times10^{34}$ & 15085.4 & 1.83 & 8.43 \\
NGC\char`_3766 & 22.9 & 2427.6 & $4.08\times10^{-8}$ & $7.94\times10^{34}$ & 35431.2 & 1.36 & 14.35 \\
NGC\char`_4103 & 20.9 & 1483.2 & $2.99\times10^{-8}$ & $5.96\times10^{34}$ & 21495.8 & 1.39 & 12.33 \\
NGC\char`_4463 & 28.8 & 304.5 & $3.24\times10^{-9}$ & $5.97\times10^{33}$ & 4524.3 & 1.37 & 13.22 \\
NGC\char`_457 & 20.9 & 2938.2 & $5.93\times10^{-8}$ & $1.18\times10^{35}$ & 42582.2 & 1.37 & 14.66 \\
NGC\char`_4755 & 12.0 & 3290.4 & $2.12\times10^{-7}$ & $4.81\times10^{35}$ & 45732.7 & 1.41 & 12.13 \\
NGC\char`_5606 & 18.2 & 307.6 & $8.21\times10^{-9}$ & $1.69\times10^{34}$ & 4411.5 & 1.80 & 6.80 \\
NGC\char`_581 & 27.5 & 725.0 & $8.45\times10^{-9}$ & $1.58\times10^{34}$ & 10734.0 & 2.29 & 12.97 \\
NGC\char`_6178 & 16.6 & 176.3 & $5.13\times10^{-9}$ & $1.07\times10^{34}$ & 2520.7 & 1.49 & 4.65 \\
NGC\char`_6193 & 5.1 & 1733.4 & $9.16\times10^{-7}$ & $2.65\times10^{36}$ & 22626.8 & 1.31 & 8.12 \\
NGC\char`_6231 & 13.8 & 3198.8 & $3.75\times10^{-5}$ & $4.15\times10^{37}$ & 44922.7 & 1.64 & 12.58 \\
NGC\char`_6249 & 20.4 & 336.9 & $7.12\times10^{-9}$ & $1.42\times10^{34}$ & 4874.4 & 1.34 & 3.66 \\
NGC\char`_6250 & 24.0 & 293.6 & $4.50\times10^{-9}$ & $8.67\times10^{33}$ & 4301.0 & 1.21 & 5.51 \\
NGC\char`_6318 & 9.8 & 2205.2 & $2.28\times10^{-7}$ & $5.46\times10^{35}$ & 30181.0 & 2.67 & 7.37 \\
NGC\char`_6322 & 14.8 & 325.4 & $1.34\times10^{-8}$ & $2.88\times10^{34}$ & 4593.8 & 1.50 & 7.70 \\
NGC\char`_6357 & 1.0 & 1387.8 & $1.34\times10^{-6}$ & $4.18\times10^{36}$ & 17800.2 & 3.61 & 11.86 \\
NGC\char`_6383 & 4.0 & 724.0 & $2.78\times10^{-7}$ & $7.75\times10^{35}$ & 9534.7 & 1.24 & 5.51 \\
NGC\char`_6396 & 15.1 & 899.1 & $3.52\times10^{-8}$ & $7.54\times10^{34}$ & 12714.7 & 2.84 & 6.16 \\
NGC\char`_6451 & 25.7 & 1633.0 & $2.18\times10^{-8}$ & $4.14\times10^{34}$ & 24046.7 & 2.86 & 9.63 \\
NGC\char`_6531 & 12.9 & 670.0 & $3.70\times10^{-8}$ & $8.26\times10^{34}$ & 9359.9 & 1.27 & 5.61 \\
NGC\char`_654 & 9.8 & 2547.7 & $2.63\times10^{-7}$ & $6.31\times10^{35}$ & 34868.0 & 1.67 & 13.97 \\
NGC\char`_6561 & 18.6 & 297.8 & $7.58\times10^{-9}$ & $1.55\times10^{34}$ & 4278.3 & 1.39 & 6.22 \\
NGC\char`_659 & 15.5 & 1486.7 & $5.54\times10^{-8}$ & $1.18\times10^{35}$ & 21061.4 & 2.83 & 7.58 \\
NGC\char`_6611 & 2.1 & 1697.0 & $1.98\times10^{-6}$ & $6.34\times10^{36}$ & 21637.2 & 2.28 & 8.58 \\
NGC\char`_6613 & 18.2 & 130.7 & $2.04\times10^{-9}$ & $3.94\times10^{33}$ & 1913.7 & 1.41 & 4.24 \\
NGC\char`_663 & 29.5 & 3820.9 & $3.88\times10^{-8}$ & $7.12\times10^{34}$ & 56872.5 & 2.63 & 13.39 \\
NGC\char`_6823 & 2.4 & 1369.6 & $1.30\times10^{-6}$ & $4.06\times10^{36}$ & 17574.7 & 2.19 & 7.03 \\
NGC\char`_6834 & 19.1 & 1733.4 & $4.21\times10^{-8}$ & $8.56\times10^{34}$ & 24945.5 & 1.84 & 12.87 \\
NGC\char`_6871 & 5.5 & 2822.3 & $1.25\times10^{-6}$ & $3.53\times10^{36}$ & 37026.6 & 1.63 & 10.75 \\
NGC\char`_6910 & 13.5 & 1193.8 & $5.97\times10^{-8}$ & $1.32\times10^{35}$ & 16736.5 & 2.40 & 10.93 \\
NGC\char`_6913 & 21.9 & 686.9 & $1.26\times10^{-8}$ & $2.49\times10^{34}$ & 9990.0 & 2.15 & 6.35 \\
NGC\char`_7039 & 21.9 & 215.2 & $3.96\times10^{-9}$ & $7.80\times10^{33}$ & 3129.6 & 1.07 & 5.12 \\
NGC\char`_7067 & 6.2 & 822.1 & $2.68\times10^{-7}$ & $7.34\times10^{35}$ & 10876.5 & 2.94 & 9.51 \\
NGC\char`_7128 & 26.3 & 1236.5 & $1.58\times10^{-8}$ & $2.98\times10^{34}$ & 18240.8 & 2.38 & 13.43 \\
NGC\char`_7129 & 21.4 & 95.6 & $7.59\times10^{-10}$ & $1.35\times10^{33}$ & 1436.9 & 1.61 & 2.60 \\
NGC\char`_7160 & 14.5 & 211.8 & $9.05\times10^{-9}$ & $1.96\times10^{34}$ & 2986.1 & 1.51 & 5.82 \\
NGC\char`_7235 & 9.3 & 1551.8 & $1.79\times10^{-7}$ & $4.34\times10^{35}$ & 21165.0 & 2.36 & 15.22 \\
NGC\char`_7261 & 11.2 & 706.7 & $5.31\times10^{-8}$ & $1.23\times10^{35}$ & 9772.1 & 2.50 & 9.54 \\
NGC\char`_7380 & 4.2 & 810.7 & $3.74\times10^{-7}$ & $1.07\times10^{36}$ & 10622.2 & 2.20 & 8.85 \\
NGC\char`_7510 & 19.5 & 2567.7 & $5.96\times10^{-8}$ & $1.20\times10^{35}$ & 37016.4 & 1.90 & 12.33 \\
NGC\char`_7788 & 17.4 & 941.2 & $2.76\times10^{-8}$ & $5.73\times10^{34}$ & 13450.3 & 1.49 & 5.75 \\
NGC\char`_869 & 15.1 & 3678.7 & $1.44\times10^{-7}$ & $3.09\times10^{35}$ & 52022.7 & 2.31 & 14.41 \\
NGC\char`_884 & 17.8 & 2293.3 & $6.42\times10^{-8}$ & $1.32\times10^{35}$ & 32828.8 & 2.31 & 7.91 \\
Negueruela\char`_1 & 17.0 & 421.7 & $1.30\times10^{-8}$ & $2.71\times10^{34}$ & 6015.2 & 2.08 & 14.98 \\
Patchick\char`_94 & 3.5 & 3692.3 & $4.75\times10^{-6}$ & $1.54\times10^{37}$ & 46930.1 & 4.84 & 5.19 \\
Pismis\char`_1 & 29.5 & 540.1 & $5.49\times10^{-9}$ & $1.01\times10^{34}$ & 8039.5 & 1.91 & 5.38 \\
Pismis\char`_11 & 3.3 & 590.5 & $1.60\times10^{-7}$ & $4.29\times10^{35}$ & 7852.2 & 2.64 & 8.19 \\
Pismis\char`_17 & 25.1 & 143.3 & $2.00\times10^{-9}$ & $3.82\times10^{33}$ & 2106.2 & 1.57 & 2.92 \\
Pismis\char`_27 & 5.4 & 212.4 & $9.18\times10^{-9}$ & $1.99\times10^{34}$ & 2992.5 & 1.67 & 5.89 \\
Pismis\char`_5 & 10.0 & 276.1 & $1.88\times10^{-8}$ & $4.30\times10^{34}$ & 3830.4 & 1.78 & 6.22 \\
Pismis\char`_Moreno\char`_1 & 8.9 & 162.2 & $3.99\times10^{-9}$ & $8.11\times10^{33}$ & 2333.4 & 1.28 & 7.52 \\
Pozzo\char`_1 & 9.5 & 342.7 & $3.45\times10^{-8}$ & $8.25\times10^{34}$ & 4693.5 & 0.42 & 5.91 \\
RSG\char`_8 & 26.9 & 204.6 & $2.50\times10^{-9}$ & $4.68\times10^{33}$ & 3023.1 & 0.82 & 5.78 \\
Riddle\char`_4 & 26.9 & 392.1 & $4.78\times10^{-9}$ & $8.97\times10^{33}$ & 5794.1 & 2.20 & 12.13 \\
Roslund\char`_2 & 11.5 & 730.1 & $5.21\times10^{-8}$ & $1.20\times10^{35}$ & 10112.5 & 2.11 & 9.91 \\
Roslund\char`_4 & 17.0 & 314.9 & $9.69\times10^{-9}$ & $2.02\times10^{34}$ & 4492.8 & 2.35 & 6.82 \\
Ruprecht\char`_127 & 13.5 & 715.9 & $3.58\times10^{-8}$ & $7.90\times10^{34}$ & 10037.1 & 2.42 & 8.05 \\
Ruprecht\char`_138 & 9.5 & 1086.9 & $1.19\times10^{-7}$ & $2.86\times10^{35}$ & 14849.9 & 2.60 & 4.42 \\
Ruprecht\char`_32 & 5.0 & 257.0 & $1.55\times10^{-8}$ & $3.49\times10^{34}$ & 3579.5 & 2.76 & 5.03 \\
Ruprecht\char`_44 & 14.5 & 731.1 & $1.20\times10^{-5}$ & $9.23\times10^{36}$ & 10303.4 & 2.58 & 7.25 \\
Ruprecht\char`_47 & 29.5 & 334.6 & $3.40\times10^{-9}$ & $6.23\times10^{33}$ & 4979.7 & 1.46 & 3.57 \\
Ruprecht\char`_65 & 22.9 & 237.2 & $3.98\times10^{-9}$ & $7.76\times10^{33}$ & 3461.8 & 1.55 & 5.61 \\
Ruprecht\char`_71 & 21.4 & 229.7 & $4.43\times10^{-9}$ & $8.76\times10^{33}$ & 3334.6 & 1.34 & 3.92 \\
Ruprecht\char`_77 & 12.3 & 192.0 & $6.73\times10^{-9}$ & $1.43\times10^{34}$ & 2725.4 & 2.28 & 4.91 \\
Ruprecht\char`_78 & 17.8 & 1250.3 & $3.50\times10^{-8}$ & $7.22\times10^{34}$ & 17898.9 & 3.11 & 4.70 \\
Ruprecht\char`_94 & 24.5 & 351.2 & $5.14\times10^{-9}$ & $9.85\times10^{33}$ & 5152.8 & 1.70 & 6.40 \\
SAI\char`_113 & 8.9 & 560.8 & $7.22\times10^{-8}$ & $1.77\times10^{35}$ & 7623.0 & 2.06 & 6.32 \\
SAI\char`_24 & 6.9 & 627.1 & $1.52\times10^{-7}$ & $4.01\times10^{35}$ & 8367.3 & 1.93 & 10.29 \\
SAI\char`_25 & 1.6 & 490.5 & $9.58\times10^{-8}$ & $2.47\times10^{35}$ & 6584.9 & 3.20 & 4.79 \\
Schuster\char`_1 & 6.3 & 84.5 & $5.15\times10^{-10}$ & $8.81\times10^{32}$ & 1282.4 & 2.66 & 5.59 \\
Sher\char`_1 & 13.2 & 702.4 & $1.20\times10^{-5}$ & $9.22\times10^{36}$ & 9830.3 & 3.56 & 14.92 \\
Shorlin\char`_1 & 3.2 & 211.2 & $9.08\times10^{-9}$ & $1.97\times10^{34}$ & 2976.8 & 3.15 & 3.45 \\
Sigma\char`_Ori & 1.3 & 113.4 & $1.34\times10^{-9}$ & $2.50\times10^{33}$ & 1677.5 & 1.16 & 4.58 \\
Skiff\char`_J2330+60.2 & 20.4 & 354.4 & $7.49\times10^{-9}$ & $1.50\times10^{34}$ & 5127.5 & 2.72 & 7.75 \\
Stephenson\char`_1 & 28.2 & 89.5 & $6.11\times10^{-10}$ & $1.06\times10^{33}$ & 1352.5 & 0.41 & 5.57 \\
Stock\char`_16 & 6.0 & 340.2 & $3.40\times10^{-8}$ & $8.10\times10^{34}$ & 4661.1 & 2.59 & 6.78 \\
Stock\char`_17 & 10.0 & 80.1 & $4.25\times10^{-10}$ & $7.11\times10^{32}$ & 1221.9 & 1.80 & 2.59 \\
Stock\char`_18 & 13.2 & 245.3 & $1.29\times10^{-8}$ & $2.86\times10^{34}$ & 3432.4 & 2.07 & 5.43 \\
Stock\char`_20 & 19.1 & 254.1 & $6.17\times10^{-9}$ & $1.25\times10^{34}$ & 3656.3 & 2.27 & 8.50 \\
Stock\char`_8 & 14.5 & 1778.7 & $7.67\times10^{-8}$ & $1.66\times10^{35}$ & 25066.0 & 1.59 & 10.28 \\
Teutsch\char`_1 & 27.5 & 378.2 & $4.41\times10^{-9}$ & $8.22\times10^{33}$ & 5599.7 & 2.29 & 3.88 \\
Teutsch\char`_125 & 8.7 & 247.3 & $1.39\times10^{-8}$ & $3.10\times10^{34}$ & 3452.6 & 2.41 & 4.35 \\
Teutsch\char`_127 & 11.5 & 352.9 & $2.52\times10^{-8}$ & $5.78\times10^{34}$ & 4887.5 & 2.42 & 6.40 \\
Teutsch\char`_13 & 7.9 & 252.2 & $1.47\times10^{-8}$ & $3.29\times10^{34}$ & 3517.8 & 2.46 & 4.46 \\
Teutsch\char`_23 & 5.6 & 537.3 & $1.23\times10^{-7}$ & $3.22\times10^{35}$ & 7180.1 & 3.18 & 5.13 \\
Teutsch\char`_30 & 28.8 & 217.4 & $2.31\times10^{-9}$ & $4.26\times10^{33}$ & 3230.6 & 2.04 & 9.60 \\
Teutsch\char`_42 & 1.3 & 480.0 & $9.02\times10^{-8}$ & $2.31\times10^{35}$ & 6451.7 & 6.39 & 7.12 \\
Teutsch\char`_8 & 4.0 & 236.6 & $1.24\times10^{-8}$ & $2.74\times10^{34}$ & 3311.6 & 2.04 & 4.79 \\
Teutsch\char`_85 & 4.0 & 760.3 & $3.17\times10^{-7}$ & $8.91\times10^{35}$ & 9990.8 & 3.29 & 14.30 \\
Trumpler\char`_1 & 28.8 & 572.4 & $6.09\times10^{-9}$ & $1.12\times10^{34}$ & 8504.6 & 2.43 & 5.24 \\
Trumpler\char`_11 & 27.5 & 437.2 & $5.10\times10^{-9}$ & $9.50\times10^{33}$ & 6473.1 & 1.42 & 5.12 \\
Trumpler\char`_15 & 8.9 & 925.4 & $1.19\times10^{-7}$ & $2.92\times10^{35}$ & 12578.7 & 2.03 & 7.16 \\
Trumpler\char`_16 & 13.5 & 1106.9 & $1.20\times10^{-5}$ & $9.21\times10^{36}$ & 15518.1 & 2.64 & 7.19 \\
Trumpler\char`_17 & 20.0 & 1001.2 & $2.22\times10^{-8}$ & $4.46\times10^{34}$ & 14459.0 & 1.46 & 6.59 \\
UBC\char`_101 & 28.2 & 525.0 & $5.84\times10^{-9}$ & $1.08\times10^{34}$ & 7785.7 & 2.36 & 5.47 \\
UBC\char`_103 & 22.4 & 883.0 & $1.55\times10^{-8}$ & $3.04\times10^{34}$ & 12865.6 & 2.69 & 5.51 \\
UBC\char`_10a & 14.1 & 205.4 & $8.31\times10^{-9}$ & $1.79\times10^{34}$ & 2901.5 & 1.39 & 5.77 \\
UBC\char`_126 & 20.0 & 434.4 & $9.62\times10^{-9}$ & $1.93\times10^{34}$ & 6273.1 & 1.89 & 4.37 \\
UBC\char`_130 & 27.5 & 605.8 & $7.06\times10^{-9}$ & $1.32\times10^{34}$ & 8969.1 & 1.79 & 13.10 \\
UBC\char`_135 & 24.0 & 654.5 & $1.00\times10^{-8}$ & $1.93\times10^{34}$ & 9586.0 & 2.66 & 12.33 \\
UBC\char`_159 & 14.5 & 112.1 & $1.26\times10^{-9}$ & $2.34\times10^{33}$ & 1662.3 & 1.02 & 4.10 \\
UBC\char`_167 & 21.9 & 224.6 & $4.13\times10^{-9}$ & $8.14\times10^{33}$ & 3266.6 & 1.42 & 3.50 \\
UBC\char`_173 & 13.2 & 2024.1 & $1.06\times10^{-7}$ & $2.36\times10^{35}$ & 28327.4 & 2.47 & 6.03 \\
UBC\char`_174 & 10.0 & 1724.5 & $1.69\times10^{-7}$ & $4.02\times10^{35}$ & 23641.9 & 2.41 & 7.31 \\
UBC\char`_176 & 6.9 & 1082.9 & $2.62\times10^{-7}$ & $6.92\times10^{35}$ & 14448.0 & 1.88 & 5.23 \\
UBC\char`_177 & 27.5 & 184.0 & $2.14\times10^{-9}$ & $4.00\times10^{33}$ & 2724.1 & 1.21 & 6.29 \\
UBC\char`_178 & 10.2 & 307.9 & $2.56\times10^{-8}$ & $5.97\times10^{34}$ & 4244.0 & 1.90 & 7.69 \\
UBC\char`_17a & 18.6 & 228.3 & $5.81\times10^{-9}$ & $1.19\times10^{34}$ & 3279.6 & 1.00 & 4.28 \\
UBC\char`_17b & 11.5 & 173.7 & $4.91\times10^{-9}$ & $1.02\times10^{34}$ & 2485.0 & 0.95 & 3.65 \\
UBC\char`_182 & 13.5 & 151.5 & $3.22\times10^{-9}$ & $6.45\times10^{33}$ & 2191.4 & 1.24 & 2.52 \\
UBC\char`_186 & 23.4 & 3003.5 & $4.82\times10^{-8}$ & $9.33\times10^{34}$ & 43915.4 & 1.37 & 12.97 \\
UBC\char`_187 & 17.8 & 115.6 & $1.38\times10^{-9}$ & $2.59\times10^{33}$ & 1709.5 & 1.07 & 2.58 \\
UBC\char`_188 & 26.3 & 363.0 & $4.63\times10^{-9}$ & $8.74\times10^{33}$ & 5354.4 & 1.39 & 4.60 \\
UBC\char`_19 & 6.9 & 142.8 & $2.71\times10^{-9}$ & $5.36\times10^{33}$ & 2075.0 & 1.99 & 3.05 \\
UBC\char`_190 & 18.6 & 624.7 & $1.59\times10^{-8}$ & $3.25\times10^{34}$ & 8973.9 & 2.14 & 5.78 \\
UBC\char`_191 & 18.2 & 677.6 & $1.81\times10^{-8}$ & $3.72\times10^{34}$ & 9717.3 & 2.27 & 14.64 \\
UBC\char`_204 & 14.8 & 322.6 & $1.32\times10^{-8}$ & $2.86\times10^{34}$ & 4553.9 & 1.70 & 3.96 \\
UBC\char`_207 & 15.1 & 122.7 & $1.68\times10^{-9}$ & $3.19\times10^{33}$ & 1805.2 & 0.88 & 1.35 \\
UBC\char`_224 & 14.8 & 1135.3 & $4.66\times10^{-8}$ & $1.01\times10^{35}$ & 16026.3 & 1.47 & 13.12 \\
UBC\char`_236 & 18.2 & 477.8 & $1.28\times10^{-8}$ & $2.62\times10^{34}$ & 6851.7 & 2.85 & 5.24 \\
UBC\char`_238 & 19.5 & 1395.5 & $3.24\times10^{-8}$ & $6.54\times10^{34}$ & 20117.6 & 2.50 & 15.16 \\
UBC\char`_261 & 18.6 & 861.4 & $2.19\times10^{-8}$ & $4.48\times10^{34}$ & 12374.7 & 1.73 & 4.74 \\
UBC\char`_263 & 25.1 & 384.6 & $5.38\times10^{-9}$ & $1.03\times10^{34}$ & 5653.5 & 1.49 & 6.17 \\
UBC\char`_266 & 12.9 & 949.5 & $5.25\times10^{-8}$ & $1.17\times10^{35}$ & 13266.1 & 1.69 & 8.83 \\
UBC\char`_267 & 12.9 & 695.0 & $3.84\times10^{-8}$ & $8.57\times10^{34}$ & 9709.4 & 1.71 & 8.02 \\
UBC\char`_287 & 23.4 & 513.9 & $8.24\times10^{-9}$ & $1.60\times10^{34}$ & 7513.2 & 1.67 & 5.67 \\
UBC\char`_291 & 25.1 & 416.8 & $5.83\times10^{-9}$ & $1.11\times10^{34}$ & 6127.2 & 1.75 & 4.91 \\
UBC\char`_293 & 20.0 & 648.4 & $1.44\times10^{-8}$ & $2.89\times10^{34}$ & 9364.3 & 1.67 & 4.88 \\
UBC\char`_31 & 26.3 & 278.7 & $3.56\times10^{-9}$ & $6.71\times10^{33}$ & 4111.6 & 1.53 & 2.55 \\
UBC\char`_318 & 11.2 & 1944.4 & $1.46\times10^{-7}$ & $3.38\times10^{35}$ & 26886.0 & 1.84 & 7.65 \\
UBC\char`_320 & 17.8 & 1135.3 & $3.18\times10^{-8}$ & $6.56\times10^{34}$ & 16252.9 & 2.43 & 5.61 \\
UBC\char`_323 & 8.9 & 1694.8 & $2.18\times10^{-7}$ & $5.36\times10^{35}$ & 23037.2 & 1.54 & 12.74 \\
UBC\char`_334 & 20.0 & 2375.6 & $5.26\times10^{-8}$ & $1.06\times10^{35}$ & 34307.6 & 2.56 & 5.57 \\
UBC\char`_341 & 21.9 & 992.8 & $1.83\times10^{-8}$ & $3.60\times10^{34}$ & 14438.6 & 2.24 & 5.08 \\
UBC\char`_342 & 14.1 & 464.8 & $2.11\times10^{-8}$ & $4.59\times10^{34}$ & 6539.0 & 1.66 & 5.58 \\
UBC\char`_343 & 20.0 & 762.3 & $1.69\times10^{-8}$ & $3.39\times10^{34}$ & 11009.2 & 1.61 & 5.05 \\
UBC\char`_344 & 3.5 & 3715.1 & $5.02\times10^{-6}$ & $1.65\times10^{37}$ & 47141.1 & 2.63 & 10.01 \\
UBC\char`_345 & 14.5 & 477.1 & $2.06\times10^{-8}$ & $4.46\times10^{34}$ & 6724.1 & 2.55 & 8.34 \\
UBC\char`_352 & 6.9 & 1227.5 & $2.97\times10^{-7}$ & $7.85\times10^{35}$ & 16376.7 & 2.74 & 4.56 \\
UBC\char`_353 & 29.5 & 416.7 & $4.23\times10^{-9}$ & $7.76\times10^{33}$ & 6202.9 & 2.17 & 3.61 \\
UBC\char`_355 & 24.5 & 675.6 & $9.89\times10^{-9}$ & $1.90\times10^{34}$ & 9913.6 & 2.53 & 4.65 \\
UBC\char`_357 & 6.6 & 1270.1 & $3.46\times10^{-7}$ & $9.27\times10^{35}$ & 16888.0 & 2.63 & 5.02 \\
UBC\char`_367 & 21.9 & 269.3 & $4.96\times10^{-9}$ & $9.76\times10^{33}$ & 3916.2 & 2.50 & 7.15 \\
UBC\char`_368 & 22.4 & 245.9 & $4.32\times10^{-9}$ & $8.46\times10^{33}$ & 3582.6 & 1.49 & 4.80 \\
UBC\char`_379 & 18.2 & 431.0 & $1.15\times10^{-8}$ & $2.36\times10^{34}$ & 6181.1 & 2.50 & 4.30 \\
UBC\char`_385 & 19.5 & 141.6 & $2.61\times10^{-9}$ & $5.14\times10^{33}$ & 2059.4 & 1.50 & 4.74 \\
UBC\char`_386 & 6.5 & 153.9 & $3.41\times10^{-9}$ & $6.85\times10^{33}$ & 2222.5 & 1.65 & 6.13 \\
UBC\char`_388 & 15.5 & 1002.4 & $3.73\times10^{-8}$ & $7.97\times10^{34}$ & 14200.5 & 2.17 & 10.36 \\
UBC\char`_393 & 8.5 & 1116.9 & $1.61\times10^{-7}$ & $4.00\times10^{35}$ & 15130.2 & 1.76 & 6.15 \\
UBC\char`_412 & 21.4 & 533.9 & $1.03\times10^{-8}$ & $2.04\times10^{34}$ & 7751.1 & 2.64 & 4.51 \\
UBC\char`_414 & 29.5 & 215.1 & $2.19\times10^{-9}$ & $4.01\times10^{33}$ & 3202.1 & 1.59 & 6.11 \\
UBC\char`_415 & 23.4 & 57.1 & $1.20\times10^{-10}$ & $1.65\times10^{32}$ & 901.2 & 1.11 & 6.95 \\
UBC\char`_418 & 22.9 & 1361.2 & $2.29\times10^{-8}$ & $4.45\times10^{34}$ & 19867.3 & 2.54 & 8.10 \\
UBC\char`_423 & 16.6 & 207.3 & $6.69\times10^{-9}$ & $1.40\times10^{34}$ & 2951.8 & 2.09 & 4.67 \\
UBC\char`_438 & 4.4 & 525.3 & $1.15\times10^{-7}$ & $3.01\times10^{35}$ & 7027.3 & 2.85 & 5.54 \\
UBC\char`_455 & 13.2 & 158.9 & $3.73\times10^{-9}$ & $7.55\times10^{33}$ & 2289.9 & 2.02 & 5.32 \\
UBC\char`_464 & 14.1 & 456.5 & $2.07\times10^{-8}$ & $4.51\times10^{34}$ & 6422.5 & 2.34 & 9.44 \\
UBC\char`_468 & 11.0 & 462.3 & $3.66\times10^{-8}$ & $8.51\times10^{34}$ & 6381.1 & 2.33 & 7.06 \\
UBC\char`_479 & 20.4 & 247.7 & $5.24\times10^{-9}$ & $1.05\times10^{34}$ & 3583.6 & 1.44 & 2.80 \\
UBC\char`_482 & 26.9 & 498.6 & $6.08\times10^{-9}$ & $1.14\times10^{34}$ & 7368.5 & 1.38 & 5.12 \\
UBC\char`_487 & 26.3 & 205.2 & $2.62\times10^{-9}$ & $4.94\times10^{33}$ & 3026.8 & 1.66 & 13.36 \\
UBC\char`_489 & 10.5 & 622.7 & $5.48\times10^{-8}$ & $1.29\times10^{35}$ & 8566.5 & 1.60 & 5.90 \\
UBC\char`_504 & 11.2 & 294.1 & $2.21\times10^{-8}$ & $5.10\times10^{34}$ & 4066.2 & 1.54 & 3.74 \\
UBC\char`_51 & 10.0 & 193.2 & $6.88\times10^{-9}$ & $1.46\times10^{34}$ & 2740.9 & 1.36 & 3.10 \\
UBC\char`_510 & 15.5 & 252.1 & $9.39\times10^{-9}$ & $2.00\times10^{34}$ & 3571.6 & 1.72 & 3.92 \\
UBC\char`_515 & 8.1 & 284.2 & $2.04\times10^{-8}$ & $4.70\times10^{34}$ & 3935.3 & 1.81 & 3.62 \\
UBC\char`_517 & 22.4 & 413.5 & $7.27\times10^{-9}$ & $1.42\times10^{34}$ & 6023.8 & 1.56 & 3.13 \\
UBC\char`_522 & 15.1 & 512.2 & $2.00\times10^{-8}$ & $4.30\times10^{34}$ & 7243.7 & 1.93 & 5.36 \\
UBC\char`_525 & 11.7 & 413.8 & $2.80\times10^{-8}$ & $6.40\times10^{34}$ & 5742.0 & 1.72 & 7.54 \\
UBC\char`_526 & 18.2 & 143.8 & $2.74\times10^{-9}$ & $5.41\times10^{33}$ & 2088.5 & 1.95 & 5.64 \\
UBC\char`_535 & 17.8 & 305.0 & $8.53\times10^{-9}$ & $1.76\times10^{34}$ & 4366.0 & 1.82 & 3.80 \\
UBC\char`_541 & 21.9 & 860.1 & $1.58\times10^{-8}$ & $3.12\times10^{34}$ & 12509.4 & 2.91 & 4.96 \\
UBC\char`_547 & 15.1 & 1897.3 & $7.42\times10^{-8}$ & $1.59\times10^{35}$ & 26830.2 & 2.47 & 6.61 \\
UBC\char`_552 & 11.2 & 515.2 & $3.87\times10^{-8}$ & $8.94\times10^{34}$ & 7124.2 & 2.68 & 7.02 \\
UBC\char`_553 & 13.2 & 255.8 & $1.34\times10^{-8}$ & $2.98\times10^{34}$ & 3579.9 & 1.84 & 6.77 \\
UBC\char`_558 & 7.2 & 2244.1 & $4.83\times10^{-7}$ & $1.26\times10^{36}$ & 30041.9 & 2.56 & 5.45 \\
UBC\char`_559 & 11.5 & 1460.6 & $1.04\times10^{-7}$ & $2.39\times10^{35}$ & 20231.6 & 2.58 & 9.08 \\
UBC\char`_562 & 11.7 & 815.5 & $5.53\times10^{-8}$ & $1.26\times10^{35}$ & 11315.5 & 2.71 & 6.35 \\
UBC\char`_564 & 25.7 & 702.7 & $9.39\times10^{-9}$ & $1.78\times10^{34}$ & 10347.3 & 2.82 & 9.88 \\
UBC\char`_565 & 9.1 & 1223.7 & $1.49\times10^{-7}$ & $3.64\times10^{35}$ & 16662.4 & 2.83 & 5.93 \\
UBC\char`_568 & 5.6 & 766.9 & $3.19\times10^{-7}$ & $8.97\times10^{35}$ & 10078.7 & 3.32 & 12.28 \\
UBC\char`_578 & 21.9 & 272.9 & $5.02\times10^{-9}$ & $9.89\times10^{33}$ & 3969.7 & 2.63 & 4.20 \\
UBC\char`_584 & 14.1 & 144.9 & $2.81\times10^{-9}$ & $5.57\times10^{33}$ & 2103.0 & 1.97 & 3.89 \\
UBC\char`_587 & 14.5 & 604.3 & $2.61\times10^{-8}$ & $5.65\times10^{34}$ & 8516.5 & 3.13 & 6.07 \\
UBC\char`_590 & 11.7 & 259.7 & $1.59\times10^{-8}$ & $3.58\times10^{34}$ & 3616.6 & 2.54 & 4.66 \\
UBC\char`_604 & 15.5 & 393.0 & $1.46\times10^{-8}$ & $3.12\times10^{34}$ & 5567.0 & 2.52 & 5.27 \\
UBC\char`_608 & 20.4 & 701.9 & $1.48\times10^{-8}$ & $2.97\times10^{34}$ & 10154.4 & 2.48 & 6.44 \\
UBC\char`_609 & 5.6 & 2886.6 & $1.20\times10^{-6}$ & $3.38\times10^{36}$ & 37933.3 & 2.17 & 7.89 \\
UBC\char`_612 & 16.2 & 152.7 & $3.29\times10^{-9}$ & $6.60\times10^{33}$ & 2207.1 & 2.48 & 12.64 \\
UBC\char`_633 & 13.8 & 227.9 & $1.08\times10^{-8}$ & $2.38\times10^{34}$ & 3199.9 & 3.36 & 8.04 \\
UBC\char`_646 & 22.9 & 165.9 & $2.79\times10^{-9}$ & $5.42\times10^{33}$ & 2420.9 & 2.13 & 12.59 \\
UBC\char`_652 & 10.0 & 123.0 & $1.70\times10^{-9}$ & $3.23\times10^{33}$ & 1808.6 & 1.79 & 5.62 \\
UBC\char`_654 & 8.9 & 1458.8 & $1.88\times10^{-7}$ & $4.61\times10^{35}$ & 19830.0 & 2.63 & 4.65 \\
UBC\char`_659 & 27.5 & 209.6 & $2.44\times10^{-9}$ & $4.56\times10^{33}$ & 3103.7 & 1.96 & 4.13 \\
UBC\char`_672 & 10.5 & 287.7 & $2.11\times10^{-8}$ & $4.86\times10^{34}$ & 3981.0 & 2.00 & 2.53 \\
UBC\char`_80 & 29.5 & 219.4 & $2.23\times10^{-9}$ & $4.09\times10^{33}$ & 3265.2 & 1.42 & 2.65 \\
UBC\char`_9 & 26.3 & 33.5 & $1.77\times10^{-11}$ & $1.32\times10^{31}$ & 560.5 & 0.71 & 3.77 \\
UPK\char`_126 & 13.5 & 55.4 & $1.09\times10^{-10}$ & $1.47\times10^{32}$ & 877.5 & 1.47 & 2.65 \\
UPK\char`_150 & 25.7 & 353.8 & $4.73\times10^{-9}$ & $8.96\times10^{33}$ & 5209.9 & 1.54 & 3.72 \\
UPK\char`_166 & 26.9 & 434.2 & $5.30\times10^{-9}$ & $9.93\times10^{33}$ & 6416.4 & 0.89 & 4.57 \\
UPK\char`_169 & 13.8 & 255.1 & $1.21\times10^{-8}$ & $2.66\times10^{34}$ & 3582.2 & 1.53 & 7.93 \\
UPK\char`_201 & 15.8 & 130.8 & $2.05\times10^{-9}$ & $3.95\times10^{33}$ & 1914.0 & 1.73 & 4.31 \\
UPK\char`_38 & 14.5 & 57.7 & $1.27\times10^{-10}$ & $1.76\times10^{32}$ & 910.3 & 1.40 & 6.49 \\
UPK\char`_385 & 11.7 & 19.8 & $2.67\times10^{-12}$ & $1.02\times10^{30}$ & 354.3 & 0.36 & 2.00 \\
UPK\char`_398 & 10.7 & 56.8 & $1.20\times10^{-10}$ & $1.66\times10^{32}$ & 897.5 & 1.03 & 1.96 \\
UPK\char`_402 & 7.2 & 27.9 & $9.27\times10^{-12}$ & $5.51\times10^{30}$ & 478.2 & 1.08 & 1.81 \\
UPK\char`_422 & 19.5 & 94.7 & $7.38\times10^{-10}$ & $1.31\times10^{33}$ & 1424.8 & 0.39 & 5.45 \\
UPK\char`_436 & 25.7 & 337.2 & $4.51\times10^{-9}$ & $8.54\times10^{33}$ & 4965.4 & 1.29 & 2.97 \\
UPK\char`_445 & 12.9 & 116.6 & $1.43\times10^{-9}$ & $2.68\times10^{33}$ & 1722.2 & 1.10 & 3.20 \\
UPK\char`_599 & 19.5 & 151.9 & $3.24\times10^{-9}$ & $6.48\times10^{33}$ & 2197.3 & 0.71 & 3.68 \\
UPK\char`_607 & 15.5 & 118.0 & $1.48\times10^{-9}$ & $2.78\times10^{33}$ & 1741.3 & 1.13 & 2.66 \\
UPK\char`_621 & 26.3 & 149.9 & $1.91\times10^{-9}$ & $3.61\times10^{33}$ & 2211.6 & 0.87 & 2.92 \\
UPK\char`_640 & 25.1 & 413.5 & $5.78\times10^{-9}$ & $1.10\times10^{34}$ & 6078.2 & 0.35 & 3.62 \\
Waterloo\char`_1 & 16.6 & 633.6 & $2.04\times10^{-8}$ & $4.29\times10^{34}$ & 9023.4 & 2.40 & 6.29 \\
Westerlund\char`_1 & 7.9 & 22227.1 & $3.01\times10^{-4}$ & $3.15\times10^{38}$ & 299575.2 & 5.51 & 6.67 \\
Westerlund\char`_2 & 4.0 & 2172.3 & $2.61\times10^{-5}$ & $1.75\times10^{37}$ & 27841.0 & 3.26 & 9.19 \\
vdBergh\char`_80 & 6.5 & 128.9 & $1.97\times10^{-9}$ & $3.80\times10^{33}$ & 1887.3 & 1.35 & 5.61 \\
vdBergh\char`_85 & 4.4 & 127.0 & $1.89\times10^{-9}$ & $3.64\times10^{33}$ & 1861.6 & 1.72 & 3.72 \\
vdBergh\char`_92 & 7.8 & 363.3 & $4.08\times10^{-8}$ & $9.86\times10^{34}$ & 4959.4 & 1.22 & 6.56 \\
\hline\hline
\end{longtable}

\end{appendix}

\end{document}